\documentclass{aastex}

\usepackage{emulateapj5}
\usepackage{graphics}

\def\ni{\noindent}                                            %No indent%
\def\etal{{et\thinspace al.}\ }               %et al. NON-ITALIC for ApJ%

\newcommand{\ET}[1]{\times 10^{#1}}                  %times power of ten%

\newcommand{\ov}[1]{\overline{#1}}                             %overline%
\newcommand{\zp}[1]{\left( {#1} \right)}                    %parentheses%

\slugcomment{Final version - Cid@JHU - June/25/2000}

\shorttitle{Quasar variability and Poissonian models}

\shortauthors{R. Cid Fernandes, L. Sodr\'e \& L. Vieira}

\begin{document}

\title{Quasar Variability in the Framework of Poissonian Models}

\author{R. Cid Fernandes\altaffilmark{1}}
\affil{Department of Physics \& Astronomy, Johns Hopkins University,
Baltimore, MD, 21218}
\email{cid@pha.jhu.edu}

\author{L. Sodr\'e Jr.}
\affil{Instituto Astr\^onomico e Geof\'{\i}sico da USP, Av.\ Miguel
Stefano 4200, 04301-904 S\~ao Paulo, SP, Brazil}
\email{laerte@iagusp.usp.br}

\and

\author{L. Vieira da Silva Jr.} 
\affil{Departamento de F\'{\i}sica - CFM, Universidade Federal de Santa
Catarina, C.P 476, 88040-900 Florian\'opolis, SC, Brazil}
\email{lande@fsc.ufsc.br}

\altaffiltext{1}{Gemini Fellow. On leave of absence from Depto.\ de
F\'{\i}sica - CFM, UFSC, Florian\'opolis, SC, Brazil}

%\date{\today}

\begin{abstract}

Quasar optical monitoring campaigns are reaching a new standard of
quality as a result of long term, accurate CCD observations.  In this
paper we review the basic Poissonian formulation of quasar
variability, using it as a mathematical tool to extract relevant
parameters such as the energy, rate and lifetimes of the flares
through the analysis of observed light curves.  It is shown that in
this very general framework the well established anti-correlation
between variability amplitude and wavelength can only be understood as
an effect of an underlying spectral component which remains stable on
long time-scales, and is redder than the variable component. The
formalism is applied to the B and R light curves of 42 PG quasars
collected by the Wise Observatory group (Giveon \etal 1999).
Variability indices for these data are obtained with a Structure
Function analysis. The mean number of living flares is constrained to
be in the range between ${\cal N} \sim 5$ and 100, while their rates
are found to be of order $\nu \sim 1$--100 yr$^{-1}$. Monochromatic
optical flare energies $E_\lambda \sim 10^{46-48}$ erg$\,$\AA$^{-1}$
and life-times $\tau$ of $\sim 0.5$ to 3 yr are derived. Lower limits
of typically 25\% are established for the contribution of a
non-variable component in the R band. The substantial diversity in
these properties among quasars invalidates simple versions of the
Poissonian model in which flare energies, lifetimes and the background
contribution are treated as universal invariants.  Light Curve
simulations confirm the applicability of the method. Few significant
correlations between variability indices and multi-wavelength
properties of the quasars exist, confirming the results of Giveon
\etal The good correlation between the EW(H$\beta$) and the long term
variability amplitude is interpreted in a scenario where only the
variable component participates in the ionization of the line emitting
gas.  This idea is consistent with the observed trends of the
variability amplitude with $\lambda$, EW(HeII) and the X-ray to
optical spectral index. The parameter estimates derived under the
framework of Poissonian models are applicable to several scenarios for
the nature of quasar variability, and can help guiding, testing and
discriminating between detailed physical models.
\end{abstract}


\section{Introduction}

\label{sec:Introduction}

Variability is one of the defining properties of Active Galactic
Nuclei (AGN) and a potentially powerful discriminant among different
scenarios for the physics of the central engine. Optical--UV
variability studies in the past decade have put a lot of effort into
investigating the effects of continuum variations upon the emission
lines, leading to substantial progress in the diagnostics of the
properties of the Broad Line Region (Netzer \& Peterson 1997). Yet, no
comparable progress has been achieved in understanding the origin of
the continuum fluctuations.

In this paper we discuss quasar variability in the context of
Poissonian models. By Poissonian we mean any physical system in which
the variations are due to the stochastic superposition of independent
flares, occurring at a given mean rate but randomly distributed in
time, a process also called ``christmas-tree'', ``shot-noise'',
``discrete events'' or ``subunits'' model. This approach has two key
advantages: (1) It provides a simple mathematical framework which can
be applied to observed light curves to constrain the most relevant
flare properties, such as energy, time-scale and rate (e.g., Cid
Fernandes, Aretxaga \& Terlevich 1996). (2) It encompasses a large
class of physical models for AGN. For instance, scenarios as diverse
as accretion-disk instabilities of several kinds (Haardt, Maraschi \&
Ghisellini 1994; Kawaguchi \etal 1998), disruption of stars in the
gravitational field of super-massive black-hole (Peterson \& Ferland
1986; Ayal, Livio \& Piran 2000), stellar collisions (Keenan 1978;
Courvoisier, Paltani \& Walter 1996), supernovae (Aretxaga \&
Terlevich 1994; Aretxaga, Cid Fernandes \& Terlevich 1997), and even
extrinsic variability models such as micro-lensing (Hawkings 2000) all
share a common Poissonian nature. It is the combination of these two
points that makes the Poissonian interpretation of AGN variability
attractive, as its application to observed data may provide
feasibility tests on different theories.

The early realization that most luminous sources tend to vary less
(Uomoto, Wills \& Wils 1976; Pica \& Smith 1983) qualitatively
favored a Poissonian interpretation. However, these and subsequent
quasar monitoring studies found that the slope ($\beta$) of the
fractional variability ($\delta \equiv \sigma(L) / \ov{L}$) versus
mean luminosity relation, $\delta \propto \ov{L}^\beta$, is shallower
than the $\beta = -1/2$ value predicted in a simple Poissonian model
(Cristiani, Vio \& Andreani 1990; Trevese \etal 1994; Hook \etal 1994,
Cristiani \etal 1996, Paltani \& Courvoisier 1997), which lead some
of these studies to rule out such models. Some studies, however, find
slopes consistent with $\beta = -1/2$ (Cid Fernandes \etal 1996;
Garcia \etal 1999), while others even question the very existence of a
correlation (Bonoli \etal 1979; Netzer \& Sheffer 1983; Giallongo,
Trevese \& Vagnetti 1991; Lloyd 1984; Cimatti, Zamorani \& Marano
1993; Netzer \etal 1996).

The slope of the variability-luminosity relation has thus been a
controversial issue, with conflicting results reported in the
literature.  Selection effects and different coverages of the
luminosity-redshift plane are likely causes for these discrepancies,
as discussed in Hook \etal (1994). Garcia \etal (1999) argued that,
 although the observed variability-luminosity relation in their sample
is shallower than $\beta = -1/2$, good agreement between data and
model is achieved after taking into account the increase in
variability with frequency.  Moreover, Cid Fernandes \etal (1996)
argued that the photometric (at least for photographic data) and
sampling uncertainties, along with wavelength/redshift effects,
propagate to a poorly defined variability-luminosity
relation. Furthermore, they showed that a slope of -1/2 is only
expected in the simplest of Poissonian models, in which the flare
energy, time-scale and background contribution are held fixed as
universal constants for all objects. Discarding Poissonian models on
the basis of $\beta \ne -1/2$ is thus both risky and an oversimplified
interpretation of Poissonian processes.

Quasar variability studies are reaching a new level of quality due to
the efforts of several groups which engaged into long term,
differential CCD photometric and spectroscopic monitoring programs
(Borgeest \& Schramm 1994; Netzer \etal 1996; Sirola \etal 1998;
Giveon \etal 1999---hereinafter G99; Garcia \etal 1999; Kaspi \etal
2000). These studies represent an enormous improvement over
pre-existing variability databases, both in sampling and photometric
quality. While most previous studies (by necessity) based their
analysis on properties of the {\it ensemble} of objects, these new
data allow the study of quasar variability properties on an
object-by-object basis. With the continuation and steady improvement
of these programs, a clearer picture of the phenomenology of quasar
variability will eventually emerge.

In this paper we review the formal relations between observable
variability properties and the parameters in a general Poissonian
model (\S\ref{sec:Formalism}). This approach is here seen as a valid
step towards a physical understanding of the nature of AGN
variability, which is particularly relevant given the current lack of
an accepted paradigm for this ubiquitous phenomenon. Emphasis is put
on the use of multi-wavelength data to constrain the basic model
parameters.  In \S\ref{sec:Analysis_Giveon} we apply the formalism to
the photometric (B and R) quasar monitoring data collected over the
past decade in the Wise Observatory (G99), by means of basic light
curve statistics and an analysis of the individual Structure
Functions. Simulations are used to verify the consistency of the
results.  A discussion on the use of higher moments of the data
(skewness and kurtosis) is also presented (Appendix A).  In Section
\ref{sec:Discussion} we discuss our results and present a correlation
analysis of the variability with other properties, as well as possible
interpretations in the context of AGN models. Finally, in Section
\ref{sec:Conclusions} we summarize our main results and outline
prospects of future work.

\section{Formalism}

\label{sec:Formalism}

Poissonian models for AGN variability have been studied both
qualitatively and quantitatively, and for wavelengths across the
electromagnetic spectrum, from radio (e.g., Dent 1972), to optical--UV
(e.g., Cid Fernandes \etal 1996; Garcia \etal 1999) and X-rays (e.g.,
Lehto 1989; Almaini \etal 2000). In this section we follow the
detailed formulation of this problem developed by Cid Fernandes
(1995). The resulting formulae rest upon basic probability and
time-series theory (e.g., Papoulis 1965).

In a Poissonian scenario the instantaneous monochromatic luminosity
$L_\lambda(t)$ is due to the superposition of a variable component,
$V_\lambda(t)$, and an underlying background component $C_\lambda$:

\begin{equation}
L_\lambda(t) = V_\lambda(t) + C_\lambda
\end{equation}

\ni where $V_\lambda(t)$ is made by the superpositions of flares
$l_\lambda(t)$ with random ``birth-dates'' $t_i$:

\begin{equation}
V_\lambda(t) = \sum_i l_\lambda(t - t_i)
\end{equation}

The first two moments (mean luminosity and relative variability) of the
variable component are

\begin{equation}
\label{eq:mean_V}
\ov{V_\lambda} = \nu_\lambda E_\lambda
\end{equation}

\begin{equation}
\label{eq:rms_V}
\frac{\sigma(V_\lambda)}{\ov{V_\lambda}}  = 
   \frac{1}{(\nu_\lambda \tau_\lambda)^{1/2}}
\end{equation}

In these expressions $\nu_\lambda$ is the mean rate of flares,
$E_\lambda$ is the monochromatic energy of individual flares, and
$\tau_\lambda$ is the flare ``life-time'', defined by

\begin{equation}
\label{eq:tau}
\tau_\lambda \equiv \frac{E_\lambda^2}{\int l_\lambda^2(t) dt}
\end{equation}

It can be shown that these relations also hold when one allows for the
(likely) possibility that the flares within a given object are not all
identical, i.e., when $l_\lambda(t) = l_\lambda(t,{\bf x})$, where
${\bf x}$ denotes a general set of parameters (size, density, cooling
time, \ldots). The only modifications in this more general case is
that $E_\lambda$ above means the average $E_\lambda({\bf x})$ over the
probability distribution of ${\bf x}$, and similarly for
$\tau_\lambda$.\footnote{To be precise, if $p({\bf x})$ is the
probability density of ${\bf x}$, one finds
$$
\tau_\lambda \equiv 
  \ov{E_\lambda}^2 / 
  \int \int l_\lambda^2(t,{\bf x}) p({\bf x}) d{\bf x} dt
$$}

The total mean luminosity and relative variability are both affected
by the underlying constant component, turning eqs.~(\ref{eq:mean_V})
and (\ref{eq:rms_V}) into

\begin{equation}
\label{eq:mean_L}
\ov{L_\lambda} = \nu_\lambda E_\lambda + C_\lambda
\end{equation}

\begin{equation}
\label{eq:rms_L}
\delta_\lambda \equiv
  \frac{\sigma(L_\lambda)}{\ov{L_\lambda}}  = 
  v_\lambda \frac{1}{(\nu_\lambda \tau_\lambda)^{1/2}}
\end{equation}

\ni where $v_\lambda \equiv \ov{V_\lambda} / \ov{L_\lambda}$ is the
fraction of the mean luminosity which is actually due to the variable
component. Allowance for the possible contribution of a non-variable
component is essential, though it has often been forgotten when
discussing Poissonian models.  Physically, $C_\lambda$ can be
associated with several sources. These can be either around the
nucleus and extrinsic to the variability generation process, as the
host galaxy stellar contribution, or, more interestingly, a part of
the variable continuum source which remains stable over long
time-scales, such as the non-flaring part of an accretion disk.

The Poissonian model thus involves four basic parameters: the rate
($\nu_\lambda$), energy ($E_\lambda$) and life-time ($\tau_\lambda$)
of the flares, plus $C_\lambda$. The shape of the flares may be
regarded as a further degree of freedom, but it has little effect upon
the analysis presented here. Equations (\ref{eq:mean_L}) and
(\ref{eq:rms_L}) relate these parameters to just two observables. Even
estimating $\tau_\lambda$ through a Structure Function (SF) analysis,
one has a non-closed system, with three observables and four
variables.  Higher moments of the light curve could in principle be
used as further constraints (see Appendix A), but these are so badly
affected by sampling uncertainties that presently they do not provide
useful constraints. Another piece of information that can help
constraining the model parameters is that the constant component has
to be at most as strong as the observed minimum in the light curve:

\begin{equation}
\label{eq:C_limits}
0 \le C_\lambda \le L_{\lambda,{\rm min}}
\end{equation}

It is reasonable to presume that the same flares are seen across a
narrow spectral band, such as the optical--UV. Empirical support for
this hypothesis comes from the high similarity between the continuum
light curves in different bands (e.g., Cutri \etal 1985; Krolik \etal
1991; G99). Indeed, the whole method of reverberation mapping relies
on an equivalent hypothesis, namely, that the fluctuations in the
ionizing continuum can be mapped by those of the optical--UV continuum
(Peterson 1993). We therefore may write $\nu_\lambda = \nu$ and
$\tau_\lambda = \tau$. The developments below could also be made
allowing for $\lambda$ dependent time-scales, but we shall adopt the
simpler constant $\tau_\lambda$ scenario.

A first consequence of this assumption is that one can isolate the
spectral shape of flare energy directly from the spectral behavior of
the standard deviation $\sigma_\lambda$:

\begin{equation}
\label{eq:Energy}
E_\lambda = 
  \left( \frac{\tau}{\nu} \right)^{1/2} \sigma_\lambda
  \propto \sigma_\lambda
\end{equation}

It is well established that the amplitude of the variations increase
towards shorter wavelengths (Cutri \etal 1985; Edelson, Krolik \& Pike
1990; Kinney \etal 1991; Cristiani \etal 1997; Di Clemente \etal
1996), with most sources becoming bluer as they brighten, which
immediately tells us that the variable component is blue. This fact
also has the very interesting consequence (see eq.~[\ref{eq:rms_L}]) that
$v_\lambda$ {\it must} vary with wavelength, which {\it can only be
understood invoking an underlying component}.

Another way of seeing this is to rewrite (\ref{eq:rms_L}) as
$\delta_\lambda = v_\lambda {\cal N}^{-1/2}$, where ${\cal N} = \nu
\tau$ is the mean number of living flares at any time. Since we are
assuming that the same flares are seen in different wavebands, the
fact that $\delta_\lambda$ decreases towards the red can only be made
consistent with a Poissonian scenario if the relative contribution of
the background, 

\[c_\lambda \equiv \frac{C_\lambda}{\ov{L_\lambda}} = 1 -
v_\lambda,\]

\ni increases with $\lambda$. Of course, the decomposition of
optical--UV spectra of AGN into a variable plus a constant component
has been proposed before, both on observational and on theoretical
grounds. Here we proved that the spectral behavior of $\delta_\lambda$
{\it implies} the existence of a constant source if one is to keep
within the framework of Poissonian models.

One can construct families of possible spectra for the constant
component by writing

\begin{equation}
\label{eq:Background}
C_\lambda = \ov{L_\lambda} - {\cal N}^{1/2} \sigma_\lambda
\end{equation}

The simultaneous analysis of the light curve statistics in different
wavebands thus has interesting consequences, but we still do not have
a closed system. Except for the life-time $\tau$, which can be
estimated through a structure function analysis, there is no way to
determine absolute values for $\nu$, $E_\lambda$ or
$C_\lambda$. Nonetheless, the wavelength information, coupled with the
condition that $C_\lambda$ must satisfy (\ref{eq:C_limits}), can yield
improved constraints upon ${\cal N}$.  One may combine the above
relations to obtain

\begin{equation}
\label{eq:N_limits}
\zp{ \frac{1 - \mu_\lambda}{\delta_\lambda} }^2 
  \le {\cal N} \le 
  \zp{ \frac{1}{\delta_\lambda} }^2 
\end{equation}  

\ni where $\mu_\lambda \equiv L_{\lambda,{\rm min}} / \ov{L_\lambda}$.
Both lower and upper limits can be made more stringent by considering
the whole wavelength information available. We can thus define ${\cal
N}_{\rm min}$ by using the maximum value of the lower limit above, and
conversely for ${\cal N}_{\rm max}$. This range of allowed ${\cal N}$
translates into corresponding ranges for $\nu$, $E_\lambda$ and
$C_\lambda$. Exactly how stringent $N_{\rm min}$ is depends on how
close $L_{\lambda,{\rm min}}$ gets to $C_\lambda$. As the chances of
the light curve reaching the background level decrease with the
increasing superposition of events, one expects the $N_{\rm min}$
limit to get progressively less stringent as ${\cal N}$ increases. The
quadratic dependence on the observed quantities also conspires to
broaden the range of ${\cal N}$.

This very general and straightforward formalism can be applied to
several variability data sets. In the next section, we apply it to one
of the best sets of quasar optical light curves presently available.

\section{Analysis of the Wise Observatory quasar light curves}

\label{sec:Analysis_Giveon}

\subsection{Description of the data}

\label{sec:Description_of_the_Data}

Giveon \etal (1999) have presented the results of a long term B and R
photometric monitoring of 42 optically selected nearby quasars from
the Palomar Green (PG) sample. The observations were collected with
the Wise Observatory 1m telescope over the 1991 to 1998 period.
Spectrophotometric measurements at Wise and Steward Observatories were
used to complement the light curves of 13 of the objects. Previous
results on this and related monitoring campaigns have been published
in Maoz \etal (1994), Netzer \etal (1996), and a spectroscopic study
of a subsample of the objects has been recently concluded (Kaspi \etal
2000). The reader is referred to G99 for details of the
observations. The following quantities are medians over the sample:
$n_{obs} = 32$ observations per object, rest-frame sampling interval
of 33 days, rest-frame light curve span of 5.9 years, photometric
uncertainty $= 0.015$ mag (in B), $z = 0.16$; $M_B = -23.3$.

The apparent magnitude light curves were converted to monochromatic
luminosities $L_\lambda$ at 4400 (B) and 6400 \AA\ (R) using the same
cosmological parameters as G99 ($H_0 = 70$ km$\,$s$^{-1}\,$Mpc$^{-1}$,
$q_0 = 0.2$, $\Lambda = 0$). The K-correction was also performed as in
G99, assuming a power-law spectrum, whose slope is defined by the
median B $-$ R color. This was a minor correction because of the low
redshifts in this sample. Galactic extinction corrections were made
with the values of $A_B$ extracted from NED\footnote{The NASA/IPAC
Extragalactic Database (NED) is operated by the Jet Propulsion
Laboratory, California Institute of Technology, under contract with
the National Aeronautics and Space Administration.}, and the
extinction law of Cardelli, Clayton \& Mathis (1989, with $R_V =
3.1$), but were mostly negligible ($A_B \le 0.17$, median $= 0.03$)
because of the high galactic latitude of the objects (Schmidt \& Green
1983). Emission lines contribute $\sim 10\%$ to the B and R band
luminosities of PG quasars (G99); their effect upon the present
analysis is also negligible.

The analysis presented below was carried out with the resulting
$L_B(t)$ and $L_R(t)$ light curves. We note that the estimates of
${\cal N}$, $\nu$, $\tau$ and the background fractional contribution
are cosmology, extinction and K-correction independent. Only the
absolute values of $E_\lambda$ and $C_\lambda$ depend on such factors.

\subsection{Structure Function Analysis: Estimating time-scales and 
the asymptotic variance}

\label{sec:SF_analysis}

The most commonly employed tool to extract information on the
variability time-scales out of quasar light curves is the Structure
Function (Simonetti, Cordes \& Heeschen 1985; Hook \etal 1994;
Cristiani \etal 1996).  The SF is defined by the mean of $[L(t +
\Delta t) - L(t)]^2$ over a light curve. For a Poissonian sequence of
flares the SF is known to be simply proportional to the SF of a single
isolated flare (Appendix B), and therefore only has structure for
time-scales shorter than the flare duration. For large $\Delta t$ the
SF converges to twice the intrinsic variance of the process:
$SF_\lambda(\Delta t) \rightarrow 2 \sigma_\lambda^2 = 2 E_\lambda^2
\nu / \tau$ (eq.~[\ref{eq:rms_L}]). This happens because variations on
such long time-scales effectively correspond to independent samples of
the $L_\lambda(t)$ process, as the light curve loses memory of its
past.

The SF provides estimates for both the variability time scale and
amplitude. We use the notation $\sigma^2_{\lambda,SF}$ to distinguish
the asymptotic variance derived from the SF from that computed
directly from the light curve $\sigma_\lambda^2 = \ov{L_\lambda^2} -
\ov{L_\lambda}^2 - \epsilon_\lambda^2$ (where the last term corrects
for the small effects of photometric error), which may be somewhat
underestimated for light curves not much longer than the correlation
time-scale since they do not sample the whole power contained in the
fluctuation power spectrum. Likewise, $\delta_{\lambda,SF} \equiv
\sigma_{\lambda,SF} / \ov{L_\lambda}$ denotes the long term net
variability.

The SFs were fit with theoretical functions corresponding to different
flare evolution models.  In all fits we are interested in just two
quantities: The flare life-time $\tau$, defined by (\ref{eq:tau}), and
$\sigma^2_{\lambda,{\rm SF}}$.  Four models for $l(t)$ were explored:
(1) square; (2) exponential; (3) symmetric triangle and (4) asymmetric
triangle. Equations and plots for the corresponding SFs are displayed
in Appendix B. These are clearly toy-models for the radiative
evolution of physical flares, but as shown in Appendix B
(Fig.~\ref{fig:theoretical_SFs}) the SF is not very sensitive to
$l(t)$. While it is unfortunate that little information about $l(t)$
can be retrieved with this technique, this does make the estimates of
$\tau$ and $\sigma^2_{\lambda,{\rm SF}}$ more robust.  This is
confirmed by the fits, which showed that all $l(t)$ models produce
very similar parameters (with the exception of exponential flares,
which in some cases do not converge and are generally poorer). For
this reason, we present only the results for square flares.

Let $L_i$ ($i=1, \ldots N$) be the luminosity of a quasar at an epoch
$t_i$ in a given spectral band. The SF of this light curve may be
defined as follows.  If $t_i$ and $t_j$ are two distinct epochs of the
light curve, and $t_i > t_j$, then

\begin{equation}
s_{ij} = s(\Delta t_{ij}) = (L_i - L_j)^2 - (\epsilon_i + \epsilon_j)^2
\end{equation}

\ni where $\Delta t_{ij} = (t_i-t_j)/(1+z)$ is the interval between
these two epochs in the quasar rest-frame. In this expression, the
observational errors $\epsilon$ are subtracted in quadrature from the
luminosity difference in order to ``remove'' the  observational noise
from the SF.  The data to be fitted, hence, are the set of pairs
$(\Delta t_{ij}, s_{ij})$. The SF parameters were estimated by direct
minimization of the sum of the square residuals between data and
model. Note that no binning was applied to data before the
fitting. All but the fourth model in Appendix B have two parameters: a
time scale $\tau$ and the asymptotic limit
$SF(\infty)=2\sigma_{\lambda,SF}^2$ (the fourth model in Appendix B
has two time-scales). The errors in the parameters were obtained by
bootstrap. In this technique an observed distribution with $n_{obs}$
data points is resampled many times,  each time picking a random set
of  $n_{obs}$ data points (allowing for  repetition) out of the
original ones. The underlying hypothesis is that the actually observed
values trace the distribution of the measured quantities. For each
resampled data set the parameters are fitted, and the parameter errors
are estimated from the variance of the values estimated in all
resampled data sets. Initially the B and R SFs were fitted
separately. A global estimate of $\tau$ combining the B and R data was
then performed, fitting both SFs with the same value of  $\tau$ but
different $\sigma_{B,SF}^2$ and $\sigma_{R,SF}^2$.

%***FIG***FIG***FIG***FIG***FIG***FIG***FIG***FIG***FIG***FIG***FIG***
\begin{figure*}[t]
\epsscale{1.9} 
%\plotone{fig1.eps}
\plotone{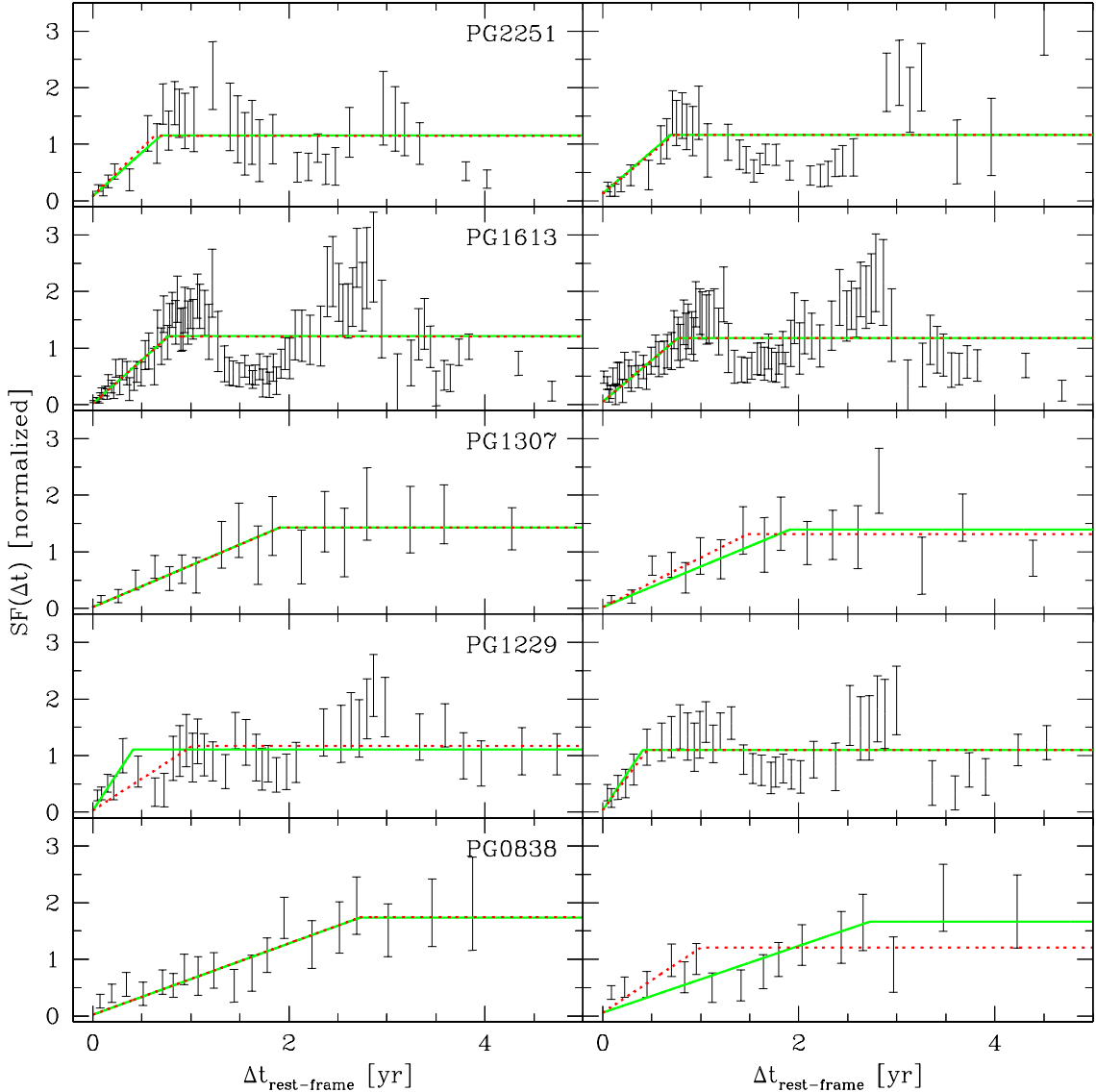}
\caption{Illustration of the Structure Functions and their fits for 5
quasars. Left and right panels show the results for B and R bands
respectively. Dotted lines correspond to the individual SF fits, while
solid lines indicate the SF resulting from the combined B and R fits.
The $(L(t+\Delta t) - L(t))^2$ differences were grouped in bins of 25
points and the error bars in the SF were computed with the bootstrap
method. This was done for plotting purposes only, since the fits do
not involve binning the SF. All fits shown correspond to square
flares. The SFs are normalized to twice the observed light curve
variance ($2 \sigma_\lambda^2$) for convenience.}
\label{fig:SF_fits}
\end{figure*}
%***FIG***FIG***FIG***FIG***FIG***FIG***FIG***FIG***FIG***FIG***FIG***

B and R SFs for five objects are illustrated in Fig.~\ref{fig:SF_fits}
along with their individual and global fits. The SFs in this figure
were computed binning the squared luminosity differences in $\Delta t$
such that each bin contained 25 points. This was done just for
plotting purposes, as the fitting procedure does not involve
binning. The error bars in Fig.~\ref{fig:SF_fits} were computed
bootstrapping the light curves 1000 times.

Though the fits were satisfactory for most objects (e.g., PG 0838 and
1307 in Fig.~\ref{fig:SF_fits}), very often the SFs exhibit complex
shapes with ups and downs (e.g., PG 1613 and 2251).  Such a non
monotonic behavior does not rule out the hypothesis that the light
curve is made up by the Poissonian superposition of flares with simple
profiles, as large departures from the predicted SF also occur for
simulated light curves when these are sampled a finite number of
times, as will be explained in Fig.~\ref{fig:SF_simulations_X_obs} and
\S\ref{sec:SF_simulations}. Theoretical SFs are derived under the
assumption of infinitely long light curves, and it would be naive to
expect observed SFs to exhibit the smooth, well behaved shapes
predicted by theory. This point has to be kept in mind when evaluating
the fits performed in this section. One must therefore exert caution
when interpreting the fit parameters, particularly $\tau$.

Results of the fits are presented for all 42 sources in Table 1, along
with other useful light curve statistics. Though in many cases the
life-times inferred from both wavelengths agreed well, we find a large
scatter around the $\tau_B = \tau_R$ line (see also G99). For the
reasons discussed above, we believe this is more likely an artifact of
the finite span of the light curves than a real effect, and in what
follows we shall only refer to the values of $\tau$ obtained with the
combined B and R fits. The asymptotic net standard deviations
$\delta_{B,SF}$ and $\delta_{R,SF}$ in Table 1 also correspond to
those obtained in the combined fits, but we remark that these were
very similar for the individual and global fits.

For four objects, PG 0026, 1100, 1427 and 1626, we find that the SF
does not converge over the time span of the observations.  All these
quasars present rising or decreasing trends over all the length of the
observations (Fig.~2 of G99). This suggests either long ($\ga 5$ yr)
time scales, an underlying slowly varying component, or else may be
indicative of a non-statistically stationary process. In addition, for
PG 1114 and 1512 we find time-scales close to the length of the light
curve. These six objects were discarded from further analysis.

\begin{figure*}[t]
\epsscale{1.5} 
%\plotone{fig2.eps}
\plotone{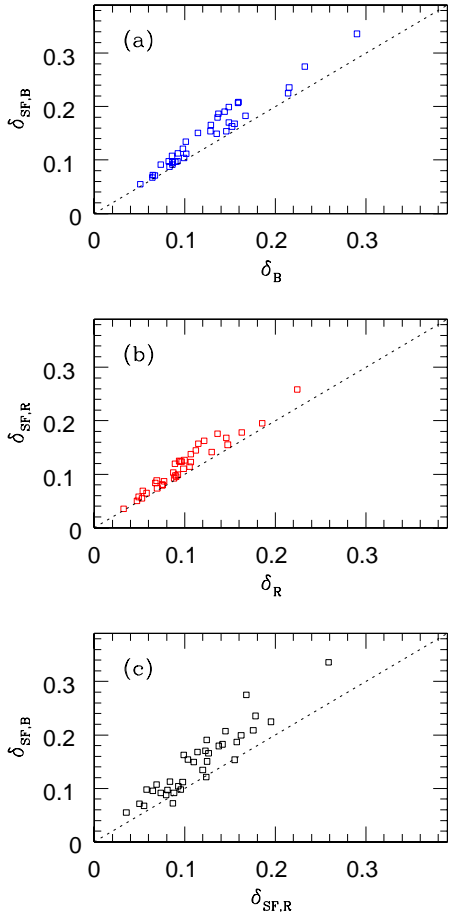}
\caption{Relation between the net variability indices computed from
the light curve, $\delta = \sigma / \ov{L}$, and those computed with
the asymptotic variance inferred from the SF. (a) B-band. (b) R band. (c) 
Comparison of B and R band net variability.}
\label{fig:rms_X_rms_SF} 
\end{figure*}

Fig.~\ref{fig:rms_X_rms_SF} shows how the net variability measured
directly from the light curve compares with the SF-determined long
term net variability.  One finds that $\delta_{\lambda,SF}$ is some
17\% larger than $\delta_\lambda$ for both B and R, confirming that
the latter slightly underestimates the long term variability. Also
shown in Fig.~\ref{fig:rms_X_rms_SF}c is a comparison of the net
variabilities in the B and R bands, to illustrate the fact that
$\delta_{B,SF} > \delta_{R,SF}$ by typically $\sim 30\%$.

%***FIG***FIG***FIG***FIG***FIG***FIG***FIG***FIG***FIG***FIG***FIG***
%***FIG***FIG***FIG***FIG***FIG***FIG***FIG***FIG***FIG***FIG***FIG***

The diversity of time-scales obtained in this analysis (from $\sim
0.5$ to 3 yr) is consistent with the results of G99, who found a wide
range of zero-crossing auto-correlation time-scales.  In fact, our
values of $\tau$ are similar to theirs, but with a substantial
scatter. In the context of Poissonian models, this diversity
reinforces the suspicion (\S1) that quasar flares are not all the
same, and that the $\delta \propto \ov{L}^{-1/2}$ may not be
universal. Indeed, in this more general scenario, the ``generalized
Poissonian model'' of Cid Fernandes \etal (1996), every quasar lies in
one of a {\it family} of $\delta \propto \ov{L}^{-1/2}$ laws (see
Fig.~\ref{fig:Correlations}). Members of each such family have
different rates, but same energy, life-time and background
contribution (eqs.~[\ref{eq:mean_L}] and [\ref{eq:rms_L}]). Also, this
intrinsic diversity casts doubts upon the meaning and usefulness of
{\it ensemble} SFs.

\begin{deluxetable}{lrrrrrrrrrrrr}
\tabletypesize{\scriptsize}
%\tabletypesize{\small}
\tablecaption{Variability Indices and Structure Function Fits}
\tablewidth{0pc}
\tablehead{
\colhead{PG}&
\colhead{$\log \ov{L_B}$}&
\colhead{$\log \ov{L_R}$}&
\colhead{$\delta_B$}&
\colhead{$\delta_{B,{\rm SF}}$}&
\colhead{$\delta_R$}&
\colhead{$\delta_{R,{\rm SF}}$}&
\colhead{$\frac{\sigma_{B,{\rm SF}}}{\sigma_{R,{\rm SF}}}$}&
\colhead{$\mu_B$}&
\colhead{$\mu_R$}&
\colhead{$\tau_B$}&
\colhead{$\tau_R$}&
\colhead{$\tau$}\cr
\colhead{(1)}&
\colhead{(2)}&
\colhead{(3)}&
\colhead{(4)}&
\colhead{(5)}&
\colhead{(6)}&
\colhead{(7)}&
\colhead{(8)}&
\colhead{(9)}&
\colhead{(10)}&
\colhead{(11)}&
\colhead{(12)}&
\colhead{(13)}}
\startdata
%% Output of NAMARRAFINAL_Tables_1and2_for_Paper.for
%% SQUare fits of Laerte used.
%% File: NAMARRAFINAL_Table1_for_Paper.out.latex_format
%% Cid@Broadview - June/19/2000
0026 & 41.37 & 41.11 & 0.18 & 0.37 & 0.14 & 0.28 & 2.31 & 0.62 & 0.63 & 5.67 & 5.72 & 5.72\tablenotemark{a} \cr
0052 & 41.53 & 40.95 & 0.22 & 0.24 & 0.16 & 0.18 & 5.04 & 0.53 & 0.62 & 0.67 & 0.70 & 0.70 \cr
0804 & 41.39 & 41.09 & 0.16 & 0.21 & 0.14 & 0.18 & 2.35 & 0.79 & 0.83 & 2.41 & 1.92 & 2.24 \cr
0838 & 40.62 & 40.93 & 0.14 & 0.19 & 0.10 & 0.12 & 0.75 & 0.71 & 0.80 & 2.73 & 1.00 & 2.73 \cr
0844 & 40.73 & 40.57 & 0.09 & 0.10 & 0.09 & 0.10 & 1.49 & 0.80 & 0.83 & 0.57 & 0.56 & 0.54 \cr
0923 & 41.45 & 41.12 & 0.15 & 0.17 & 0.11 & 0.11 & 3.14 & 0.72 & 0.81 & 0.63 & 0.66 & 0.64 \cr
0953 & 41.68 & 41.31 & 0.14 & 0.18 & 0.11 & 0.14 & 3.02 & 0.79 & 0.81 & 2.68 & 1.26 & 2.61 \cr
1001 & 40.95 & 40.60 & 0.21 & 0.22 & 0.19 & 0.20 & 2.56 & 0.78 & 0.80 & 0.34 & 0.49 & 0.36 \cr
1012 & 41.49 & 40.88 & 0.10 & 0.12 & 0.10 & 0.12 & 4.01 & 0.84 & 0.80 & 2.13 & 2.58 & 2.35 \cr
1048 & 40.65 & 40.73 & 0.23 & 0.27 & 0.15 & 0.17 & 1.35 & 0.64 & 0.69 & 1.62 & 1.48 & 1.58 \cr
1100 & 41.76 & 41.41 & 0.08 & 0.14 & 0.05 & 0.08 & 3.75 & 0.81 & 0.88 & 3.20 & 0.65 & 5.15\tablenotemark{a} \cr
1114 & 41.17 & 40.93 & 0.14 & 0.21 & 0.11 & 0.19 & 1.97 & 0.70 & 0.74 & 2.37 & 6.79 & 4.05\tablenotemark{a} \cr
1115 & 40.99 & 40.80 & 0.15 & 0.17 & 0.11 & 0.12 & 2.16 & 0.76 & 0.77 & 1.34 & 1.27 & 1.34 \cr
1121 & 41.32 & 41.06 & 0.13 & 0.17 & 0.10 & 0.13 & 2.41 & 0.69 & 0.79 & 2.15 & 1.77 & 2.08 \cr
1151 & 41.18 & 40.84 & 0.14 & 0.15 & 0.10 & 0.11 & 3.00 & 0.79 & 0.84 & 1.05 & 1.07 & 1.05 \cr
1202 & 41.08 & 40.72 & 0.29 & 0.34 & 0.22 & 0.26 & 3.01 & 0.52 & 0.60 & 1.19 & 0.97 & 1.10 \cr
1211 & 41.25 & 40.95 & 0.14 & 0.19 & 0.11 & 0.16 & 2.35 & 0.76 & 0.77 & 3.39 & 2.39 & 3.31 \cr
1226 & 42.36 & 41.98 & 0.10 & 0.13 & 0.09 & 0.12 & 2.68 & 0.80 & 0.79 & 2.20 & 2.64 & 2.35 \cr
1229 & 40.55 & 40.35 & 0.15 & 0.15 & 0.15 & 0.16 & 1.56 & 0.73 & 0.75 & 1.02 & 0.45 & 0.41 \cr
1307 & 41.28 & 40.97 & 0.13 & 0.15 & 0.09 & 0.10 & 3.05 & 0.75 & 0.85 & 1.92 & 1.48 & 1.91 \cr
1309 & 41.48 & 41.21 & 0.08 & 0.10 & 0.05 & 0.06 & 3.09 & 0.85 & 0.90 & 1.90 & 1.55 & 1.87 \cr
1322 & 41.19 & 40.88 & 0.07 & 0.07 & 0.05 & 0.05 & 2.89 & 0.88 & 0.90 & 0.56 & 0.73 & 0.56 \cr
1351 & 41.15 & 40.91 & 0.10 & 0.10 & 0.09 & 0.09 & 1.95 & 0.83 & 0.82 & 0.44 & 1.19 & 0.46 \cr
1354 & 41.18 & 40.86 & 0.15 & 0.16 & 0.09 & 0.10 & 3.51 & 0.73 & 0.86 & 0.42 & 0.53 & 0.47 \cr
1402 & 41.31 & 40.93 & 0.08 & 0.09 & 0.08 & 0.08 & 2.61 & 0.85 & 0.87 & 0.37 & 0.33 & 0.36 \cr
1404 & 40.56 & 40.43 & 0.09 & 0.09 & 0.07 & 0.07 & 1.71 & 0.85 & 0.89 & 0.47 & 0.44 & 0.47 \cr
1411 & 41.03 & 40.80 & 0.09 & 0.11 & 0.07 & 0.08 & 2.31 & 0.80 & 0.88 & 1.82 & 2.29 & 2.20 \cr
1415 & 40.77 & 40.59 & 0.06 & 0.07 & 0.05 & 0.06 & 1.85 & 0.90 & 0.89 & 0.35 & 0.28 & 0.34 \cr
1426 & 40.78 & 40.62 & 0.16 & 0.21 & 0.11 & 0.14 & 2.08 & 0.65 & 0.77 & 3.12 & 3.20 & 3.12 \cr
1427 & 41.08 & 40.74 & 0.20 & 0.40 & 0.16 & 0.32 & 2.79 & 0.61 & 0.75 & 6.32 & 5.94 & 6.05\tablenotemark{a} \cr
1444 & 41.06 & 40.85 & 0.06 & 0.07 & 0.08 & 0.09 & 1.36 & 0.88 & 0.84 & 0.87 & 1.31 & 1.04 \cr
1512 & 41.67 & 41.31 & 0.11 & 0.19 & 0.09 & 0.16 & 2.63 & 0.78 & 0.80 & 4.54 & 4.54 & 4.54\tablenotemark{a} \cr
1519 & 40.89 & 40.67 & 0.09 & 0.10 & 0.06 & 0.06 & 2.42 & 0.83 & 0.88 & 0.79 & 1.42 & 1.08 \cr
1545 & 41.52 & 41.15 & 0.15 & 0.20 & 0.12 & 0.16 & 2.91 & 0.72 & 0.79 & 2.56 & 2.60 & 2.62 \cr
1613 & 41.41 & 41.12 & 0.10 & 0.11 & 0.09 & 0.10 & 2.23 & 0.79 & 0.77 & 0.77 & 0.77 & 0.77 \cr
1617 & 41.04 & 40.63 & 0.17 & 0.18 & 0.13 & 0.14 & 3.32 & 0.68 & 0.77 & 0.66 & 0.79 & 0.69 \cr
1626 & 41.04 & 40.79 & 0.29 & 0.64 & 0.24 & 0.55 & 2.07 & 0.38 & 0.45 & 5.59 & 6.11 & 7.67\tablenotemark{a} \cr
1700 & 41.97 & 41.77 & 0.07 & 0.09 & 0.07 & 0.09 & 1.65 & 0.87 & 0.90 & 1.75 & 1.79 & 1.75 \cr
1704 & 42.06 & 41.83 & 0.11 & 0.15 & 0.09 & 0.12 & 2.03 & 0.79 & 0.83 & 2.06 & 2.07 & 2.06 \cr
2130 & 40.90 & 40.56 & 0.09 & 0.10 & 0.08 & 0.08 & 2.62 & 0.81 & 0.83 & 0.69 & 1.78 & 0.93 \cr
2233 & 41.52 & 41.00 & 0.09 & 0.11 & 0.05 & 0.07 & 5.20 & 0.81 & 0.87 & 1.65 & 3.09 & 2.20 \cr
2251 & 41.82 & 41.54 & 0.05 & 0.05 & 0.03 & 0.04 & 2.93 & 0.89 & 0.93 & 0.63 & 0.73 & 0.70 \cr
\enddata
\tablenotetext{a}{SF did not converge.}
\tablecomments{(1) quasar name (only first four PG numbers)\\
(2,3) mean B and R monochromatic luminosities 
(in erg$\,$s$^{-1}\,$\AA$^{-1}$)\\
(4,6) net variability $\delta$ in B and R, as computed directly
from the light curves\\
(5,7) asymptotic net variability in B and R, as computed from the SF analysis\\
(8) ratio of asymptotic net variabilities in B and R (estimate of $E_B / E_R$)\\
(9,10) $L_\lambda,{\rm min} / \ov{L_\lambda}$ at B and R\\
(11,12) B and R life-times as inferred from the individual SF fits (in yr)\\
(13) life-time obtained by the combined fit to the B and R SFs (in yr)}

\end{deluxetable}

\subsection{Constraints on the energies, rate and background contribution}

\label{sec:Constraints}

The asymptotic net variabilities $\delta_{B,SF}$ and $\delta_{R,SF}$
obtained above, along with the values of $\ov{L}_\lambda$ and
$\mu_\lambda$ (listed in Table 1), can be directly applied on
eq.~(\ref{eq:N_limits}) to constrain the allowed range of ${\cal N}$,
$c_B$ and $c_R$.  We remark that this is nearly independent of the SF
analysis, since $\delta_{\lambda,SF}$ is not too different from
$\delta_\lambda$ (Fig.~\ref{fig:rms_X_rms_SF}) and $\tau$ is not
employed in these constraints. The results of this calculation are
presented in Table 2.

Except for 3 out of 36 quasars, the upper limit ${\cal N}_{max}$ is
always given by the B band, since, as already noted by G99, the
variations there are larger than in R (Fig.~\ref{fig:rms_X_rms_SF}c),
in agreement with the general tendency of AGN. As anticipated
(\S\ref{sec:Formalism}), the ranges of the obtained ${\cal N}$s are
wide (factor of $\sim 20$). Still, they provide useful estimates of a
physically meaningful quantity in Poissonian models. From Table 2 one
concludes that the typical number of living events in the G99 quasars
is of order 5--100.

A corollary of ${\cal N}_{max}$ being defined by the B light curves is
that non trivial lower limits for the background component can only be
obtained for the R band for the majority of objects (Table 2).  In
some cases, like for PG 1309 and 1354, one finds {\it lower limits} of
as much as 40\% in R, indicating a substantial contribution from an
underlying constant spectral component.  The upper limits for $c_R$
are given by $\mu_R$, tabulated in Table 1, which for these two
quasars are 90 and 86\% respectively.

\subsubsection {Energies and rates}

The ${\cal N}$ and $c_\lambda$ limits above make no explicit use of
the life-times. Estimates for the allowed range of flare rates and
energies require knowledge of $\tau$. These are listed in Table 2 for
the life-times found in \S\ref{sec:SF_analysis}.  Rates between 1 and
a few hundred yr$^{-1}$, and monochromatic energies from $3\ET{45}$ to
$10^{49}$ erg$\,$\AA$^{-1}$ are found for $E_B$ and $E_R$. Note
that the energies for individual quasars are more constrained
(typically to within a factor of 4) than either $\nu$ or ${\cal N}$,
since they only depend on the square root of $\nu$. Furthermore,
inspection of Table 2 shows that there is not a single value of either
$E_B$ or $E_R$ that is simultaneously compatible with all lower and
upper limits for the G99 quasars. This, again, points to a {\it
diversity} of flare properties in quasars.

\begin{deluxetable}{lrrrrrrrrrr}
%\tabletypesize{\scriptsize}
\tabletypesize{\small}
\tablecaption{Estimates of Flare Energies and Rates}
\tablewidth{0pc}
\tablehead{
\colhead{PG}&
\colhead{${\cal N}_{min}$}&
\colhead{${\cal N}_{max}$}&
\colhead{$\nu_{min}$}&
\colhead{$\nu_{max}$}&
\colhead{$E_{B,min}$}&
\colhead{$E_{B,max}$}&
\colhead{$E_{R,min}$}&
\colhead{$E_{R,max}$}&
\colhead{$c_{B,min}$}&
\colhead{$c_{R,min}$}\cr
\colhead{(1)}&
\colhead{(2)}&
\colhead{(3)}&
\colhead{(4)}&
\colhead{(5)}&
\colhead{(6)}&
\colhead{(7)}&
\colhead{(8)}&
\colhead{(9)}&
\colhead{(10)}&
\colhead{(11)}}
\startdata
%% Output of NAMARRAFINAL_Tables_1and2_for_Paper.for
%% SQUare fits of Laerte used.
%% File: NAMARRAFINAL_Table2_for_Paper.out.latex_format
%% Cid@Broadview - June/19/2000
0052 &    4.5 &   18.0 &    6.5 &   25.9 &  $4.1\ET{47}$ &  $8.2\ET{47}$ &  $8.2\ET{46}$ &  $1.6\ET{47}$ & 0.00 & 0.24 \cr
0804 &    1.0 &   23.0 &    0.5 &   10.3 &  $7.5\ET{47}$ &  $3.5\ET{48}$ &  $3.2\ET{47}$ &  $1.5\ET{48}$ & 0.00 & 0.16 \cr
0838 &    2.6 &   27.4 &    1.0 &   10.0 &  $1.3\ET{47}$ &  $4.2\ET{47}$ &  $1.7\ET{47}$ &  $5.6\ET{47}$ & 0.00 & 0.35 \cr
0844 &    4.4 &  104.7 &    8.1 &  194.1 &  $8.8\ET{45}$ &  $4.3\ET{46}$ &  $5.9\ET{45}$ &  $2.9\ET{46}$ & 0.00 & 0.02 \cr
0923 &    2.9 &   35.5 &    4.5 &   55.9 &  $1.6\ET{47}$ &  $5.5\ET{47}$ &  $5.0\ET{46}$ &  $1.8\ET{47}$ & 0.00 & 0.32 \cr
0953 &    1.8 &   31.1 &    0.7 &   11.9 &  $1.3\ET{48}$ &  $5.2\ET{48}$ &  $4.2\ET{47}$ &  $1.7\ET{48}$ & 0.00 & 0.23 \cr
1001 &    1.0 &   19.8 &    2.9 &   55.6 &  $5.0\ET{46}$ &  $2.2\ET{47}$ &  $2.0\ET{46}$ &  $8.7\ET{46}$ & 0.00 & 0.13 \cr
1012 &    2.6 &   65.4 &    1.1 &   27.8 &  $3.4\ET{47}$ &  $1.7\ET{48}$ &  $8.5\ET{46}$ &  $4.3\ET{47}$ & 0.02 & 0.00 \cr
1048 &    3.5 &   13.2 &    2.2 &    8.4 &  $1.7\ET{47}$ &  $3.3\ET{47}$ &  $1.2\ET{47}$ &  $2.4\ET{47}$ & 0.00 & 0.39 \cr
1115 &    3.4 &   34.5 &    2.6 &   25.7 &  $1.2\ET{47}$ &  $3.8\ET{47}$ &  $5.5\ET{46}$ &  $1.7\ET{47}$ & 0.00 & 0.28 \cr
1121 &    3.4 &   36.4 &    1.6 &   17.5 &  $3.8\ET{47}$ &  $1.2\ET{48}$ &  $1.6\ET{47}$ &  $5.1\ET{47}$ & 0.00 & 0.24 \cr
1151 &    2.0 &   44.7 &    1.9 &   42.7 &  $1.1\ET{47}$ &  $5.3\ET{47}$ &  $3.8\ET{46}$ &  $1.8\ET{47}$ & 0.00 & 0.26 \cr
1202 &    2.4 &    8.8 &    2.2 &    8.1 &  $4.7\ET{47}$ &  $9.0\ET{47}$ &  $1.6\ET{47}$ &  $3.0\ET{47}$ & 0.00 & 0.23 \cr
1211 &    2.2 &   28.6 &    0.7 &    8.6 &  $6.5\ET{47}$ &  $2.4\ET{48}$ &  $2.8\ET{47}$ &  $1.0\ET{48}$ & 0.00 & 0.16 \cr
1226 &    3.1 &   55.2 &    1.3 &   23.5 &  $3.1\ET{48}$ &  $1.3\ET{49}$ &  $1.1\ET{48}$ &  $4.9\ET{48}$ & 0.00 & 0.11 \cr
1229 &    3.2 &   41.6 &    7.7 &  101.2 &  $1.1\ET{46}$ &  $3.9\ET{46}$ &  $7.0\ET{45}$ &  $2.5\ET{46}$ & 0.01 & 0.00 \cr
1307 &    2.7 &   42.0 &    1.4 &   22.0 &  $2.7\ET{47}$ &  $1.1\ET{48}$ &  $8.9\ET{46}$ &  $3.5\ET{47}$ & 0.00 & 0.33 \cr
1309 &    2.8 &  105.3 &    1.5 &   56.2 &  $1.7\ET{47}$ &  $1.0\ET{48}$ &  $5.4\ET{46}$ &  $3.3\ET{47}$ & 0.00 & 0.41 \cr
1322 &    4.0 &  198.0 &    7.2 &  354.5 &  $1.4\ET{46}$ &  $9.6\ET{46}$ &  $4.7\ET{45}$ &  $3.3\ET{46}$ & 0.00 & 0.30 \cr
1351 &    3.7 &   92.2 &    8.1 &  200.4 &  $2.2\ET{46}$ &  $1.1\ET{47}$ &  $1.1\ET{46}$ &  $5.7\ET{46}$ & 0.00 & 0.11 \cr
1354 &    2.7 &   37.7 &    5.8 &   80.9 &  $6.0\ET{46}$ &  $2.2\ET{47}$ &  $1.7\ET{46}$ &  $6.3\ET{46}$ & 0.00 & 0.39 \cr
1402 &    3.0 &  129.9 &    8.3 &  356.7 &  $1.8\ET{46}$ &  $1.2\ET{47}$ &  $6.9\ET{45}$ &  $4.5\ET{46}$ & 0.00 & 0.10 \cr
1404 &    2.5 &  119.0 &    5.4 &  255.8 &  $4.5\ET{45}$ &  $3.1\ET{46}$ &  $2.7\ET{45}$ &  $1.8\ET{46}$ & 0.00 & 0.20 \cr
1411 &    3.2 &   79.1 &    1.5 &   36.0 &  $9.5\ET{46}$ &  $4.7\ET{47}$ &  $4.1\ET{46}$ &  $2.0\ET{47}$ & 0.00 & 0.26 \cr
1415 &    3.9 &  221.5 &   11.5 &  657.7 &  $2.8\ET{45}$ &  $2.1\ET{46}$ &  $1.5\ET{45}$ &  $1.1\ET{46}$ & 0.00 & 0.18 \cr
1426 &    2.8 &   23.3 &    0.9 &    7.5 &  $2.5\ET{47}$ &  $7.3\ET{47}$ &  $1.2\ET{47}$ &  $3.5\ET{47}$ & 0.00 & 0.30 \cr
1444 &    3.5 &  132.5 &    3.4 &  127.6 &  $2.4\ET{46}$ &  $1.4\ET{47}$ &  $1.7\ET{46}$ &  $1.1\ET{47}$ & 0.17 & 0.00 \cr
1519 &    3.6 &  109.3 &    3.3 &  100.8 &  $2.4\ET{46}$ &  $1.3\ET{47}$ &  $9.9\ET{45}$ &  $5.5\ET{46}$ & 0.00 & 0.32 \cr
1545 &    2.0 &   25.1 &    0.8 &    9.6 &  $1.1\ET{48}$ &  $3.9\ET{48}$ &  $3.8\ET{47}$ &  $1.3\ET{48}$ & 0.00 & 0.19 \cr
1613 &    5.4 &   80.0 &    7.0 &  103.6 &  $7.9\ET{46}$ &  $3.0\ET{47}$ &  $3.5\ET{46}$ &  $1.4\ET{47}$ & 0.00 & 0.13 \cr
1617 &    3.1 &   30.0 &    4.5 &   43.5 &  $8.0\ET{46}$ &  $2.5\ET{47}$ &  $2.4\ET{46}$ &  $7.4\ET{46}$ & 0.00 & 0.22 \cr
1700 &    2.1 &  119.0 &    1.2 &   67.9 &  $4.3\ET{47}$ &  $3.2\ET{48}$ &  $2.6\ET{47}$ &  $2.0\ET{48}$ & 0.00 & 0.04 \cr
1704 &    1.9 &   44.2 &    0.9 &   21.4 &  $1.7\ET{48}$ &  $8.2\ET{48}$ &  $8.3\ET{47}$ &  $4.0\ET{48}$ & 0.00 & 0.17 \cr
2130 &    4.6 &  106.4 &    5.0 &  115.0 &  $2.2\ET{46}$ &  $1.0\ET{47}$ &  $8.3\ET{45}$ &  $4.0\ET{46}$ & 0.00 & 0.17 \cr
2233 &    3.6 &   86.6 &    1.6 &   39.3 &  $2.6\ET{47}$ &  $1.3\ET{48}$ &  $5.1\ET{46}$ &  $2.5\ET{47}$ & 0.00 & 0.36 \cr
2251 &    4.2 &  330.6 &    6.0 &  475.4 &  $4.3\ET{46}$ &  $3.9\ET{47}$ &  $1.5\ET{46}$ &  $1.3\ET{47}$ & 0.00 & 0.36 \cr
\enddata
%\tablenotete$t{}{}
\tablecomments{
(2,3) minimum and maximum number of events present at any time\\ 
(4,5) minimum and maximum event rate (in yr$^{-1}$)\\
(6--9) minimum and maximum monochromatic flare energy in B and R (in
erg$\,$\AA$^{-1}$)\\
(10,11) minimum background fraction of mean luminosity in B and R}
\end{deluxetable}

\subsection{Light curve simulations}

\label{sec:Simulations}

The estimates of the basic Poissonian parameters presented here rely
on just the mean and minimum luminosities plus the SF-based estimates
of the flare life-time and asymptotic variance, this latter quantity
being little different from the light curve variance.  The actual
light curves contain much more information, as they are defined by a
particular realization of birth-dates of the flares and their detailed
radiative evolution. Retrieving this information from these
observations is however an impossible task. We have experimented using
higher moments (skewness and kurtosis) of the light curves as further
constraints, but this proved fruitless in practice (Appendix A).

Since our estimates are based on so little information, there is no
guarantee that a superposition of flares with the inferred properties
would bear any morphological resemblance to the observed light curves.
In order to verify whether Poissonian models within the bounds defined
in \S\ref{sec:Constraints} {\it can} produce quasar-looking light
curves we have performed a series of Monte Carlo light curve
simulations. Our goal here is to broadly assess the ``morphological
compatibility'' of models and data in a qualitative way, based on a
simple visual inspection. The simulations make use of the value of
$\tau$ listed in Table 1, and vary $\nu$ within its empirically
defined limits for each quasar (Table 2). The flare energy and
background luminosity are determined self consistently through
eqs.~(\ref{eq:Energy}) and (\ref{eq:Background}). Square shaped flares
were used for consistency with the results already presented.

%***FIG***FIG***FIG***FIG***FIG***FIG***FIG***FIG***FIG***FIG***FIG***
\begin{figure*}[t]
\epsscale{2} 
%\plotone{fig3.eps}
\plotone{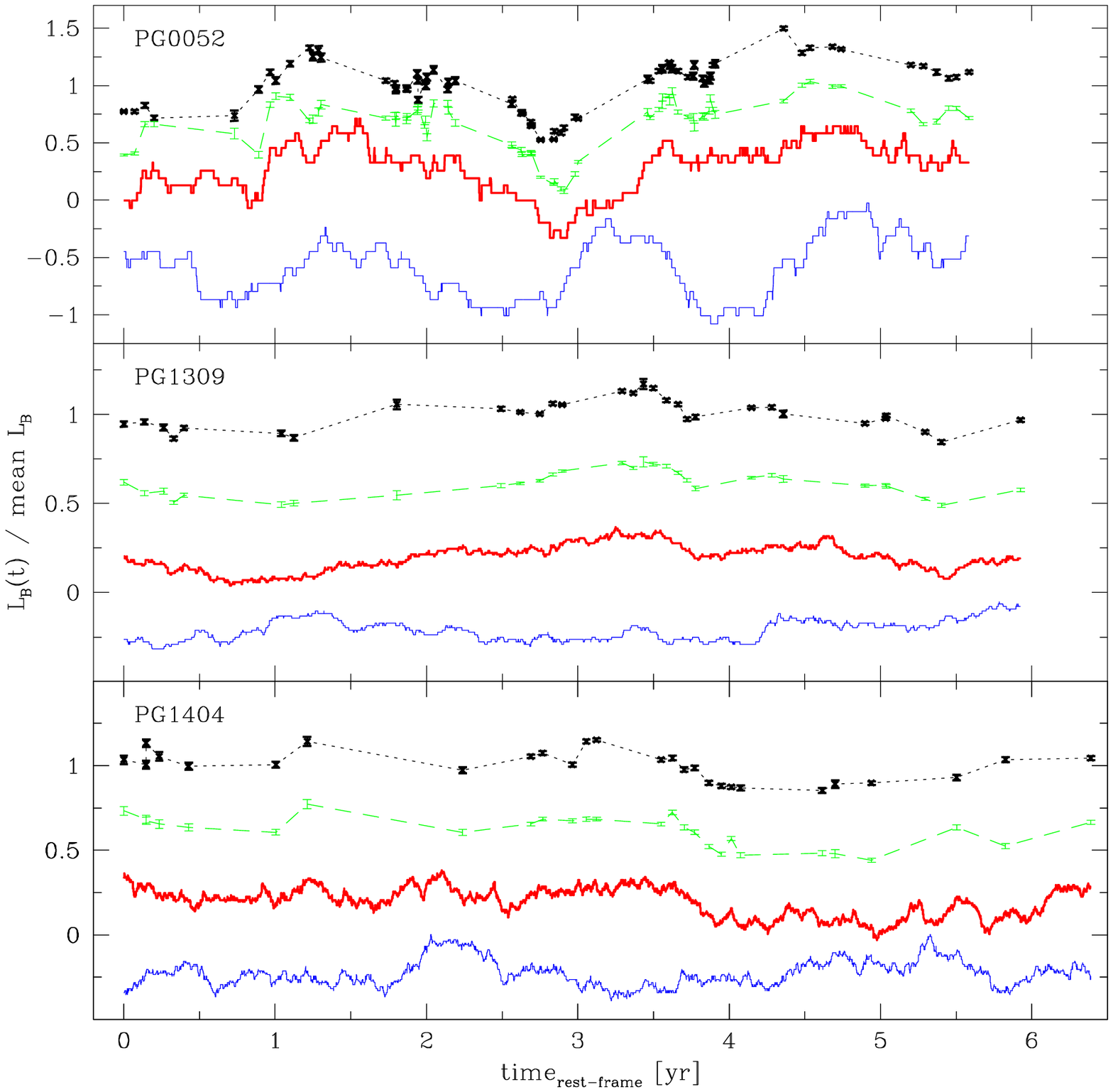}
\caption{Comparison of observed quasar B-band light curves (top dotted
lines) and simulations. Thin solid lines show a randomly chosen
simulation.  The thick lines indicate a ``best match'' model (see
text), while the dotted line show the same model after sampling as in
the data.  All light curves are normalized to $\ov{L_B}$ (Table 1),
and the simulations are shifted downwards for clarity. The global
Poissonian parameters for the simulations in thick lines are $\tau =
0.70$, 1.87 and 0.47 yr; $\nu = 19$, 47 and 223 yr$^{-1}$; $E_B =
4.8\ET{47}$, $1.8\ET{47}$ and $4.9\ET{45}$ erg$\,$\AA$^{-1}$; $c_B =
15$, 8 and 7\% for PG 0052, 1309 and 1404 respectively. The
simulations in the bottom thin lines have the same lifetimes as above,
but $\nu = 16$, 29 and 131 yr$^{-1}$; $E_B = 5.2\ET{47}$, $2.4\ET{47}$
and $6.3\ET{45}$ erg$\,$\AA$^{-1}$; $c_B = 21$, 28 and 29\% for PG
0052, 1309 and 1404 respectively. These values are all, by
construction, consistent with the limits in Table 2.}
\label{fig:LC_simulations_X_obs} 
\end{figure*}
%***FIG***FIG***FIG***FIG***FIG***FIG***FIG***FIG***FIG***FIG***FIG***

In Fig.~\ref{fig:LC_simulations_X_obs} we present three illustrative
examples of this experiment. The top curves in each panel show the B
band light curves for PG 0052, 1309 and 1404, as observed in the Wise
Observatory campaign. The thin solid line in the bottom illustrates a
randomly chosen realization of the light curve, computed with $\nu$
set to half-way between $\nu_{min}$ and $\nu_{max}$. The thick solid
line shows the simulation which, among thousands of runs, best matches
the observed light curve in a $\chi^2$ sense, whereas the dashed curve
shows the same simulation after sampling it with the same rest-frame
pattern as the data and with Gaussian perturbations added mimicking
the sequence of photometric errors reported in G99. All model light
curves were vertically displaced for clarity. By construction, their
mean luminosity is essentially identical to the observed one, as are
their variances.

Considering the simplicity of the model and the requirement of
consistency with the constraints defined in \S\ref{sec:Constraints},
the resemblance between observations and the models is very
satisfactory. The ``best-matches'' are particularly striking. Even
better matches would be obtained fitting the light curves, i.e.,
adjusting both the global parameters and the individual flare
birth-dates to optimize the residuals. Of course, the meaning and
usefulness of such detailed fits would be questionable given the
number of parameters involved. For instance, there are anywhere
between 36 and 145 flares for the length of the PG0052 light curve,
and up to 1316 in PG1404. Indeed, with so many degrees of freedom, it
might be impossible to ever disprove such models.  Unlike the flare
dates, however, one has much less liberty to play with the global
parameters, and our estimates of $\tau$, $\nu$, $E_\lambda$ and
$c_\lambda$ go a long way towards constraining the simulations to a
region of parameter space capable of producing quasar looking light
curves. Simulations with much longer or shorter time scales, for
instance, do not resemble the observations at all. We therefore
conclude that even a simple Poissonian superposition of square flares
is capable of producing light curves which are very similar to quasars
and simultaneously compatible with their basic light curve statistics.

\subsubsection{Structure Function}

\label{sec:SF_simulations}

\begin{figure*}[t]
\epsscale{2} 
%\plotone{fig4.eps}
\plotone{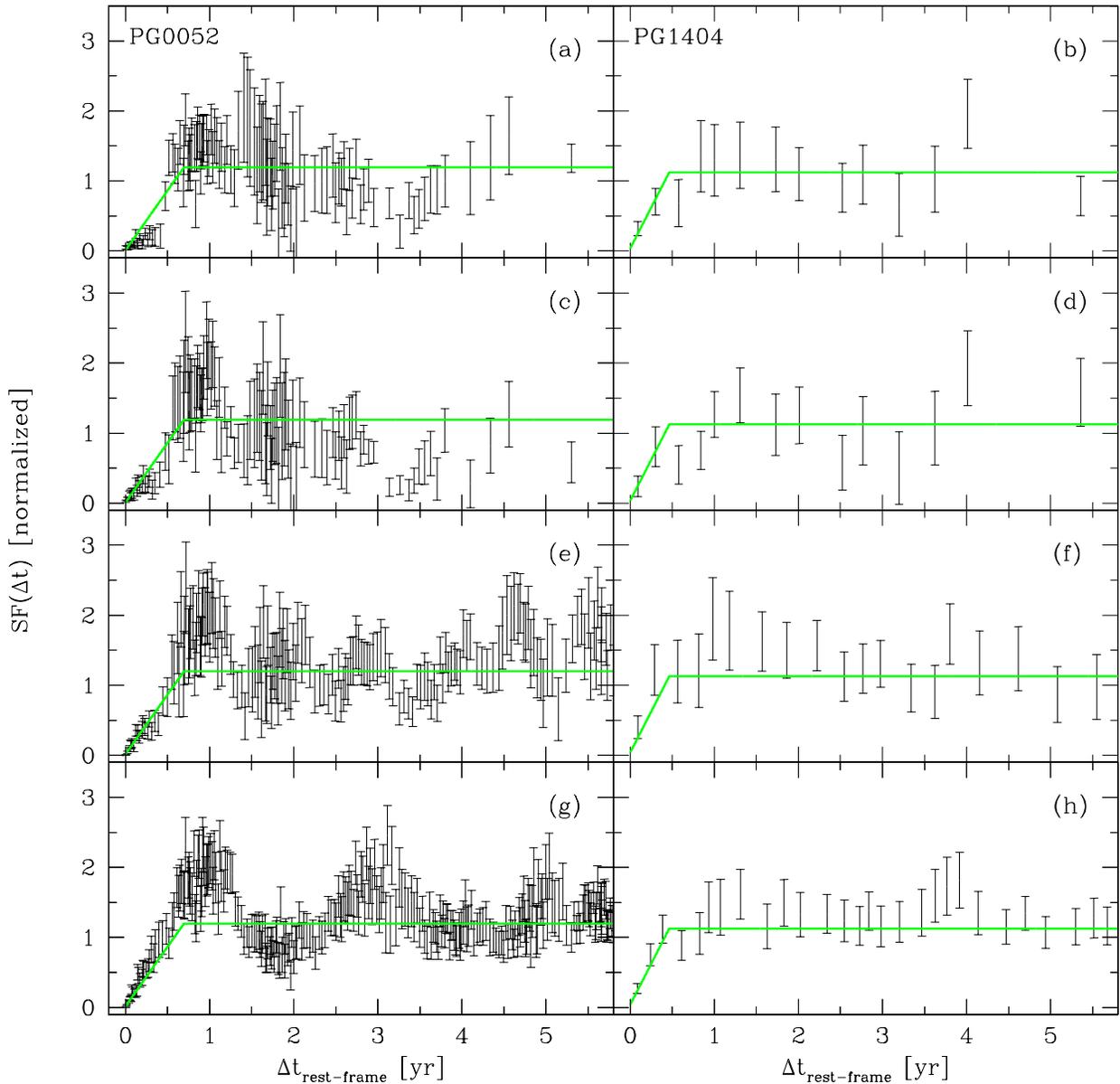}
\caption{Structure Function of PG 0052 (panel a) and PG 1404 (b) in
the B band.  Panels c and d show the SFs of the corresponding
simulated light curves (shown as dashed lines in
Fig.~\ref{fig:LC_simulations_X_obs}). Simulations twice and five times
longer than the length of the Wise Observatory monitoring of these
quasars were used to compute the SFs in panels e and f (2$\times$),
and g and h (5$\times$). In all plots the solid line marks the same
combined B and R SF fit performed upon the observed light curves. All
SFs are normalized to twice the observed $\sigma_B^2$.}
\label{fig:SF_simulations_X_obs} 
\end{figure*}

The SFs of PG 0052 and 1404 are compared to the corresponding
simulated light curves in Fig.~\ref{fig:SF_simulations_X_obs}. As
anticipated, relatively large deviations from the smooth theoretical
curve occur even for model light curves. The bottom panels show how
the SF is improved with light curves twice to five times longer than
the observed ones. These were constructed simply extending the
sampling of the model by patching together 2 (panels e and f) or 5 (g
and h) sequences of the actual observing dates for these two quasars,
thus maintaining the G99 sampling pattern. Even for such long (13--45
yr) hypothetical light curves the SF oscillates, though one sees it
gradually converging to its statistically expected shape and
amplitude. Increasing the sampling rate is not as benefic as
increasing the length of the data train, as SF oscillations on
intervals $\Delta t$ are only averaged out for $\Delta t \gg \tau$,
i.e, as independent (separated by $\sim \tau$) portions of the light
curve are sampled a statistically significant number of times. This
explains why even long light curves exhibit long term SF `noise' but
are much better defined on short time scales
(Fig.~\ref{fig:SF_simulations_X_obs}e--h).

%***FIG***FIG***FIG***FIG***FIG***FIG***FIG***FIG***FIG***FIG***FIG***
%***FIG***FIG***FIG***FIG***FIG***FIG***FIG***FIG***FIG***FIG***FIG***

These experiments illustrate the inherent difficulty in modeling SFs of
stochastic processes. As a further example, we note that the SF analysis of
a longer (11 yr) light curve of PG 1226 (= 3C 273) by Paltani, Courvoisier
\& Walter (1988) suggests a value of $\tau$ of $\sim 0.6$ yr, a factor of 4
smaller than the 2.3 yr we found for the 5.5 yr G99 monitoring of this
quasar. The SF of PG 1226 is similar to that of PG 1229 in the B band
(Fig.~\ref{fig:SF_fits}), with an initial peak at small $\Delta t$ ($\sim
0.4$ yr), followed by oscillations on longer time scales. Whereas for PG
1229 the global fits favored the small $\tau$, in PG 1226 the amplitude of
the 0.5 yr peak is smaller, and a longer $\tau$ was favored, similar to
what is seen in the R band SF of PG 0838 (Fig.~\ref{fig:SF_fits}). Even
considering the differences in technicalities of SF analysis between
different works, one is forced to conclude that the actual uncertainties in
the SF parameters is likely to substantially exceed the formal fit-errors
(about 5\% of the parameter values).

Overall, we conclude that SF analysis for individual objects is just
starting to become possible, despite the excellent quality of the G99
light curves.  We are however confident that the fits performed in
this paper are correct to within better than an order of magnitude,
and thoroughly meet our goal of providing rough but useful estimates
for the basic parameters of Poissonian models for the optical
variability of quasars.

\section{Discussion}

\label{sec:Discussion}

\begin{figure*}[t]
\epsscale{1.1} 
%\plotone{fig5.eps}
\plotone{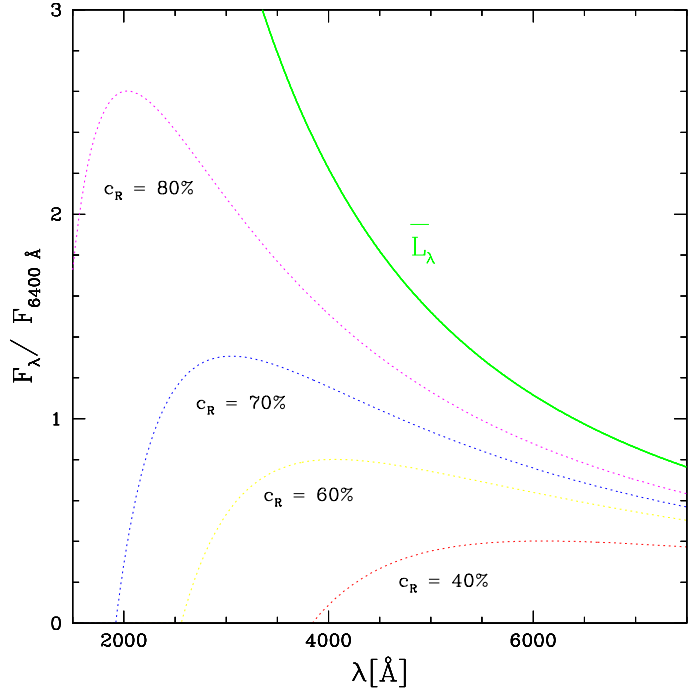}
\caption{Background components ($C_\lambda$, dotted lines) for a
hypothetical quasar with $\ov{L_\lambda} \propto \lambda^{-1.7}$ (top
solid line) and $\delta_\lambda \propto \lambda^{-1}$. The plot
illustrates how the spectral energy distribution of the constant
component can be deduced from (observationally) well determined
$\ov{L_\lambda}$ and $\delta_\lambda$, coupled with variability
derived constraints upon $c_\lambda$ at any particular wavelength.}
\label{fig:C_lambda}
\end{figure*}

Poissonian models provide a simple and elegant mathematical framework
which relates basic parameters to measurable quantities, as
demonstrated by the results presented above. The same technique can be
applied to other data sets and spectral bands, thus increasing the
number of constraints. In particular, it will be interesting to apply
it to the spectrophotometric monitoring observations of 28 quasars,
all in the G99 sample, reported in Kaspi \etal (2000). This should
yield a clearer picture of the (optical) spectral energy distributions
of the variable and constant components, as well as better constrained
parameters than was possible with only two wavelengths. An
illustration of how important it is to properly define
$\ov{L_\lambda}$ and $\delta_\lambda$ is given in
Fig.~\ref{fig:C_lambda}.

%***FIG***FIG***FIG***FIG***FIG***FIG***FIG***FIG***FIG***FIG***FIG***
%***FIG***FIG***FIG***FIG***FIG***FIG***FIG***FIG***FIG***FIG***FIG***

The $E_B$ limits listed in Table 2 encompass both the estimates of Cid
Fernandes \etal (1996) and Garcia \etal (1999) for different
samples. Their estimates, however, were based on the properties of the
{\it ensemble} of quasars, and assumed typical variability
time-scales, whereas here we attempted to treat each quasar
individually. In fact, this is the first time that constraints on
Poissonian parameters are computed on an object by object basis for a
large number of quasars.

A further comparison can be made with the results of Paltani \etal
(1988) for 3C273. We estimate from their plots a monochromatic flare
energy of $\sim 2\ET{47}$ erg$\,$\AA$^{-1}$, about an order of
magnitude smaller than $E_{B,min}$ in Table 2. This discrepancy is
explained by the already mentioned $\sim 5$ times smaller $\tau$ found
by those authors, as well as their $\sim 2$ times smaller asymptotic
net variability compared to the $\delta_{SF,B} = 0.13$ in Table 2, and
another factor of $\sim 1.5$ due to the redshift effect (see
below). The discrepancy in the asymptotic SF variance is mostly due to
the fact that they allow for a slowly ($\ga 10$ yr) varying component,
attributed to a blazar behavior.  Two of the other 6 radio loud
objects in G99 sample (PG 1100 and 1512) show evidence for long
time-scale variations which could be due to a similar slow component
(marginal evidence that the radio loud objects in this sample present
longer time-scales can be seen in Fig.~\ref{fig:Correlations}c). It is
in fact adequate to recall that AGN variability has often been
described in terms of the superposition of rapid and slowly varying
components (e.g., Lyutyi 1977, Pica \etal 1988).  A caveat in the
present analysis is thus that what we have been calling a ``constant
component'' may well correspond to an underlying slow process.

We note that no attempt was made to correct any of the variability
indices nor the inferred parameters for the variability-$\lambda$
effect which plague fixed band photometric light curves of quasars,
leading to overestimated variability amplitudes (Cristiani \etal 1996;
Cid Fernandes \etal 1996; Aretxaga \etal 1997; Garcia \etal
1999). Fortunately, the low redshifts of the G99 quasars minimize this
effect. For the median $z$ of the sample, and using the
parameterization of the $\delta \times \lambda$ relation of Garcia
\etal (1999), we estimate that the variability indices in Table 1
(columns 4 to 7) would need to be multiplied by typically 0.8 to
obtain the rest-frame indices. This would {\it increase} ${\cal
N}_{min}$, ${\cal N}_{max}$ and the corresponding $\nu$ limits by some
56\% while {\it reducing} the energies by a factor of 0.64. These are
relatively small factors considering the allowed ranges for the
parameters and were not applied in Table 2. Furthermore, such
corrections would soon become obsolete since the spectral data
collected by Kaspi \etal (2000) allows the direct computation of
rest-frame continuum variability indices. We did, however, experiment
with the $\lambda$-correction in the correlation analysis presented
next.

\subsection{Correlation Analysis}

\begin{deluxetable}{lcccccc}
%\tabletypesize{\scriptsize}
%\tabletypesize{\small}
\tablecaption{Correlation Matrix}
\tablewidth{0pc}
\tablehead{
\colhead{  }&
\colhead{$\ov{L_B}$}&
\colhead{$\ov{L_R}$}&
\colhead{$z$}&
\colhead{EW(H$\beta$)}&
\colhead{EW(HeII)}&
\colhead{$\alpha_{ox}$}}
\startdata
$\delta_{B,{\rm SF}}$ & -81.7(-37.2) & -98.3(-51.3) & -65.1(-16.8) & 0.3(0.1) & 2.3(1.2) & -1.8(-0.7)  \cr
$\delta_{R,{\rm SF}}$ & -99.3(-40.3) & -94.1(-37.6) & -58.1(-8.0)  & 0.3(0.1) & 4.4(1.1) & -1.4(-1.1)  \cr
$\tau$                &  3.9         & 0.4          &  37.8        & 7.5      & 76.6     & -0.8        \cr
$\sigma_{B,{\rm SF}} / \sigma_{R,{\rm SF}}$ 
                      &  0.1         & 8.2          & 1.9          & 33.4     & -98.5    & 18.9        \cr

$\mu_B$                & 47.0         & 67.7         &  50.9        & -0.4     & -0.5     &  0.1        \cr
\enddata
\end{deluxetable}

Perhaps the most puzzling result of G99 was the finding that there are
few convincing correlations between variability indices and a large
array of other observed properties for the PG quasars. In Table 3 we
synthesize the results of a correlation analysis analogous to that
performed by G99, but for our SF-based variability indices. The table
is similar to their Table 4, from which the multi-wavelength data was
borrowed, and lists the percentage probabilities ($P_r$) of no
correlation in a Spearman's rank test, small values indicating
significant correlations and negative values corresponding to
anticorrelations.  Entries between parentheses correspond to
correlations after correction of the variability amplitudes with the
$\delta$-$\lambda$-$z$ relation following Garcia \etal (1999).

\begin{figure*}[t]
\epsscale{2} 
%\plotone{fig6.eps}
\plotone{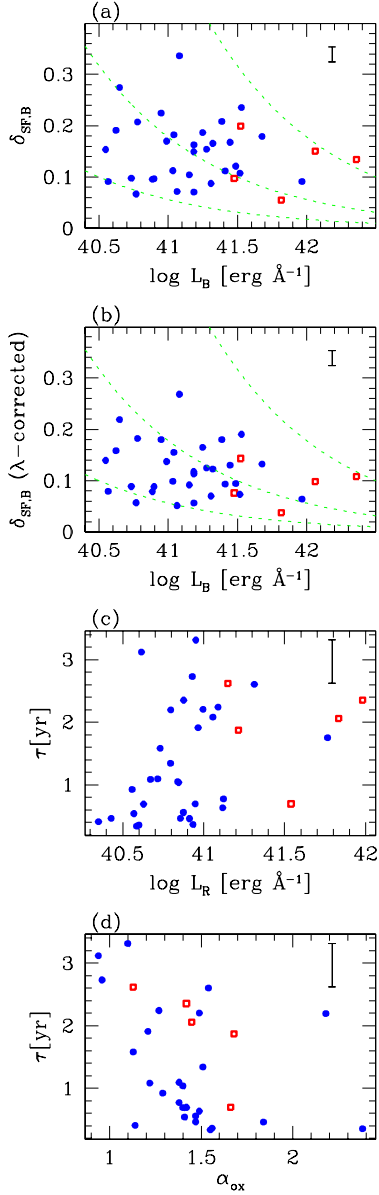}
\caption{Correlations between variability properties and luminosity
(a--c) and $\alpha_{ox}$ (d).  Filled circles mark radio quiet
objects, whereas squares correspond to radio loud sources Dotted lines
in the top panels are simple Poissonian models with $0.5 \log (1 -
c_B) E_B / \tau = 40.5$ (top line), 39.5 (middle) and 38.5
(bottom). The error bars on the top right of each panel represent the
mean formal uncertainties in $\delta_{SF}$ and $\tau$ as obtained from
the SF fits.}
\label{fig:Correlations}
\end{figure*}

The anti-correlation between the variability amplitude (here measured
by $\delta_{SF,B}$ and $\delta_{SF,R}$) and the optical luminosity
found in other studies (e.g., Hook \etal 1994) is essentially absent
for this sample, as illustrated in Fig.~\ref{fig:Correlations}a.
Somewhat more significant anti-correlations are obtained by applying
the $\lambda$ correction (Fig.~\ref{fig:Correlations}b; Table 3), but
in both cases the correlations here are even weaker than those found
by G99. The dotted lines in Fig.~\ref{fig:Correlations}a and b mark
the prediction of simple Poissonian models in which $c_\lambda$,
$E_\lambda$ and $\tau$ are identical for all quasars. This illustrates
the often claimed failure of this model (e.g. Hook \etal 1994), and
the necessity to allow for a diversity of values for these parameters
to account for the observed variability properties in the framework of
Poissonian processes.

The variability time-scale seems to increase with luminosity
(Fig.~\ref{fig:Correlations}c), but, as in G99, with a large scatter.
Previous evidence for such a correlation was reported by Cristiani
\etal (1996), in an analysis of the {\it ensemble} SF of 486 objects
spanning a much larger range of luminosities and redshifts. It is
interesting to note that applying their Model E SF-fits to the typical
absolute magnitudes of G99 quasars yield $\tau$ between $\sim 0.5$ and
2.5 yr, compatible with our individual SF fits.  We also find $\tau$
to correlate positively with the $\delta_{SF}$'s, but this may be due
to a trade-off effect in the SF fitting
(\S\ref{sec:SF_simulations}). The weak anticorrelation with
$\alpha_{ox}$ found by G99 is much stronger here
(Fig.~\ref{fig:Correlations}d).

The good correlations between the variability amplitude and the
equivalent widths of H$\beta$ and HeII$\,\lambda$ 4686 found by G99
are reproduced at about the same high level of significance
(Fig.~\ref{fig:EWHbeta}a). Unlike G99, however, we find the optical to
X-ray index $\alpha_{ox}$ to be significantly anticorrelated with both
$\delta_{SF,B}$ (Fig.~\ref{fig:EWHbeta}d) and $\delta_{SF,R}$. All
these correlations are improved with the $\lambda$-correction.

A correlation was found between the ratio of the $B$ to $R$ asymptotic
standard deviations and $\ov{L_B}$ as well as $z$. This could be due
to the redshift effects already discussed, but in this case a
$\delta$-$\lambda$ law different from that of Garcia \etal (1999)
would be implied, since the $z$-effects upon $\sigma_{B,{\rm SF}} /
\sigma_{R,{\rm SF}}$ cancels out in their
parameterization. Alternatively, if this interpretation is proved
wrong, one would conclude that the flares are bluer in more luminous
sources (eq.~\ref{eq:Energy}).

Overall, we largely confirmed G99 results, with a few minor
differences (e.g., Fig.~\ref{fig:Correlations}d) undoubtedly due to
the different methodologies employed in the definition of variability
amplitudes and time-scales. Despite the few good correlations found,
whose meaning we discuss below, a striking result of this analysis is
the large scatter present even in the best correlations obtained. In
any scenario for AGN, the variability properties must somehow be
dictated by the physical conditions prevalent in the active nucleus
(accretion rate, black hole mass, orientation, etc.). Since these
conditions must also define other aspects of AGN phenomenology, it is
natural to expect that properties driven by common causes should
exhibit some degree of correlation.  Given the controversial history
of correlations in quasar variability studies, however, it is perhaps
not so surprising that so few good correlation were found.

Regarding Poissonian models, one would obviously like to search for
correlations between the basic parameters and other properties.  The
constraints upon ${\cal N}$, $\nu$, $E_\lambda$ and $c_\lambda$
derived in this paper are however not strong enough to warrant a
proper correlation analysis. Taking the middle value between ${\cal
N}_{min}$ and ${\cal N}_{max}$ as a measure of ${\cal N}$, we obtain
an anticorrelation between ${\cal N}$ and EW(H$\beta$) and a positive
correlation with $\alpha_{ox}$, both with $P_r$ at the 1\% level. The
flare rate follows the same trends. For the reason discussed above, it
would be premature to give much emphasis to these correlations.

%***FIG***FIG***FIG***FIG***FIG***FIG***FIG***FIG***FIG***FIG***FIG***
%***FIG***FIG***FIG***FIG***FIG***FIG***FIG***FIG***FIG***FIG***FIG***

\subsubsection{Interpretation of the correlations with EW(H$\beta$)}

A simple interpretation of the significant correlation between
EW(H$\beta$) and the variability amplitude is possible by postulating
that the variable component dominates the ionization of the gas.  This
idea is also compatible with our earlier conclusion
(\S\ref{sec:Formalism}) that the variability-wavelength
anticorrelation indicates the dilution of a blue variable component by
a red underlying background. In this hypothesis both variable
($V_\lambda$) and constant ($C_\lambda$) components contribute to the
continuum under H$\beta$, but only the former is proportional to the
ionizing luminosity ($L_{ion}$), so one predicts

\begin{equation}
\label{eq:EWHbeta_1}
EW(H\beta) \propto 
	\frac{\nu E_B}{\nu E_B + C_B} 
\end{equation}

\ni where we used the B band because of its proximity to H$\beta$.
This relation may be rewritten as

\begin{equation}
\label{eq:EWHbeta_2}
EW(H\beta) \propto {\cal N}^{1/2} \delta_{SF,B} = v_B = 1 - c_B
\end{equation}

\ni which reveals a proportionality between EW(H$\beta$) and
$\delta_{SF,B}$. Obviously, the same prediction applies to EW(HeII).

\begin{figure*}[t]
\epsscale{1.8}
%\plotone{fig7.eps}
\plotone{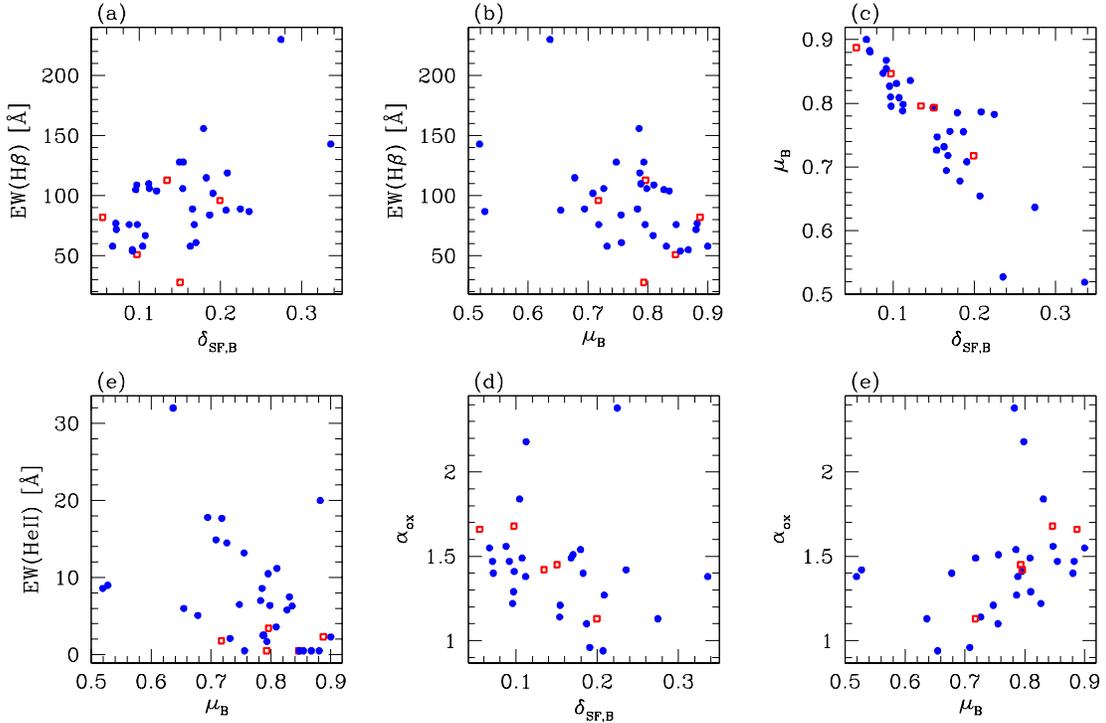}
\caption{Correlations between $\delta_{SF,B}$, $\mu_B$, EW(H$\beta$),
EW(HeII) and $\alpha_{ox}$ (the larger $\alpha_{ox}$ the softer the
spectrum). Note that $\mu_B$, the fraction of the minimum to the mean
B -band luminosity, is to be seen as an upper limit to true fraction
of the background luminosity ($c_B \le \mu_B)$.  Symbols as in
Fig.~\ref{fig:Correlations}.}
\label{fig:EWHbeta}
\end{figure*}

Using ${\cal N} = ({\cal N}_{min} + {\cal N}_{max}) / 2$, the product
${\cal N}^{1/2} \delta_{SF,B}$ correlates at the $P_r = 0.6\%$ level
with EW(H$\beta$), in agreement with the prediction. Since our
estimates of ${\cal N}$ are not independent of $\delta_{SF,B}$
(eq.~\ref{eq:N_limits}), it is perhaps more meaningful to test the
predicted relation using $c_B$, for which we have a robust upper limit
imposed by the minimum in the light curve ($\mu_B$ in Table 1). In
Fig.~\ref{fig:EWHbeta}b we see that the expected anticorrelation is
confirmed with a high significance (Table 3). Even considering that
$\mu_B$ is an upper limit to $c_B$ and the scatter in the plot, it is
interesting to see that the trend is roughly linear, as predicted.

$\mu_B$ is also strongly correlated with $\alpha_{ox}$ ($P_r =
0.1\%$), the X-ray spectral index $\alpha_x$ ($P_r = 1.8\%$), and
anticorrelated with EW([OIII]) ($P_r = 0.8\%$), [OIII] to H$\beta$
peak intensities ratio ($P_r = 0.2\%$) and EW(HeII) ($P_r = 0.5\%$).
The correlation with $\alpha_{ox}$, in particular, gives strength to
our working hypothesis, insofar as this index can be interpreted as
indicative of the ratio between the constant and variable (ionizing)
components.

By admitting that different quasars have different background
fractions, and that $V_\lambda$ dominates the ionization, we naturally
explain at least some of the correlations observed. This argues
against $C_\lambda$ being part of the ionizing source, such as hot but
non-flaring portions of the accretion disk surface. Outer, colder
regions of the disk can contribute to $C_\lambda$, as do the host
galaxy starlight and other circum-nuclear sources.

The alternative hypothesis that $L_{ion}$ is proportional to the total
$V_\lambda + C_\lambda$ components would not induce the correlations
above, while the hypothesis that $L_{ion}$ is governed by the constant
component alone can be straightforwardly rejected, since this would
not produce emission line variability.

We note that our eq.~(\ref{eq:EWHbeta_1}) is formally and
philosophically identical to eq.~(10) of Aretxaga \& Terlevich (1994),
who identify $C_\lambda$ with a young nuclear star-cluster and
$V_\lambda$ with compact SN remnants exploding out of the same cluster
(see also Aretxaga \etal 1997).  Their analysis of EW(H$\beta$) aimed
to understand the near universal value of this quantity across the AGN
luminosity scale (Binette, Fosbury \& Parker 1993).  In this model, a
range of EW(H$\beta$) values could be linked to either different SN
explosion energies, or, as is more likely, to different H$\beta$ and
continuum production efficiencies in the SN remnants (see Cid
Fernandes 1997 for a review of the pros and cons of the starburst
model).  Stochastic effects could also play a role in defining the
EW(H$\beta$) and variability relation, something which can happen in
{\it any} Poissonian scenario. Indeed, the tentatively identified
anticorrelation between ${\cal N}$ (and $\nu$) and EW(H$\beta$) may be
indicative of stochastic effects in the G99 data, and may be partly
responsible for the scatter in Fig.~\ref{fig:EWHbeta}a and d.

%***FIG***FIG***FIG***FIG***FIG***FIG***FIG***FIG***FIG***FIG***FIG***
%***FIG***FIG***FIG***FIG***FIG***FIG***FIG***FIG***FIG***FIG***FIG***

\subsection{Relation to physical models}

The ultimate goal of the Poissonian description of quasar variability
is of course to extract information on the fundamental properties of
the variability phenomenon and use them to guide and discriminate
between physical theories. A detailed discussion of the implications
of the estimates presented in \S\ref{sec:Analysis_Giveon} to different
scenarios for AGN variability is beyond the scope of the present
paper, but we would like to take one particular model as an
illustration of the possible associations between the observationally
constrained parameters discussed here and physically meaningful
quantities.

In the model proposed by Haardt \etal (1994), blobs emerge from the
surface of an accretion disk and release magnetically stored energy in
the form of rapid X-ray flares. In the X-ray regime, the background
fraction in this model would be associated with the fraction of the
disk area covered by blobs (see Figure 1 of Galeev, Rosner \& Vaiana
1979), which equals the ratio of charge and discharge times. As they
postulate the optical-UV variability to be driven by the X-ray flares,
$c_\lambda$ is related to the fraction of the reprocessed luminosity
and the optical-UV emission from the underlying stable component of
the disk, but note that comparison with the limits on $c_\lambda$
derived in this paper require allowance for the diluting effects of
larger scale sources, such as the host galaxy or an extended
scattering region. One of the predictions of this model is that at any
time there are of order 10 active blobs, independent of the source
luminosity, consistent with the limits on ${\cal N}$ established here.
Also the reprocessed luminosity, which we would identify with $(1 - c)
\ov{L} = \nu E$ in our formalism, is predicted to be similar to the
X-ray luminosity. It is not clear, however, if the time smearing of
the minutes long flares during the disk illumination which results in
the reprocessing of the blobs energies can be made compatible with the
$\sim$ 0.5--3 yr time scales implied by the SF analysis. Surely this
must impose some limits upon the geometry of the reprocessing
surface. Furthermore, it remains to be established whether the flare
energies can be made compatible with the limits laid out in
\S\ref{sec:Constraints}. A more detailed scrutiny of the Haardt \etal
(1994) model must therefore await more theoretical developments. On
the observational side, a Poissonian analysis of X-ray light curves of
quasars (similar to that done by Cid Fernandes 1995 for EXOSAT light
curves of Seyferts) could also set constraints upon this particular
model.

Despite their generality and appealing aspects, Poissonian models are
by no means the only possible description of AGN variability. For
instance, Mineshige \& Shields (1990) showed that thermal limit cycles
similar to those known to occur in dwarf novae (Warner 1995) can also
take place in AGN disks, leading to eruptions followed by quiescent
periods. Such a model, in which the degree of variability is
``self-regulated'', would invalidate the formalism employed here,
which fundamentally rests upon the hypothesis of {\it independence} of
the flares and statistical stationarity.  The same applies to
scenarios involving random walk-like or other kinds of state dependent
behavior (e.g., Begelman \& De Kool 1991; Stern, Svensson \& Sikora
1991) or periodic modulations associated to, for instance, precessing
jets (e.g., Abraham \& Romero 1999). {\it Testing} the Poissonity of
AGN light curves is however a hard task. Fourier phase coherence
studies like that performed by Krolik, Done \& Madejski (1993) for
X-ray light curves of Seyferts, or intermittency tests (Vio \etal
1992) like the one carried out by Longo \etal (1996) for the
historical light curve of NGC 4151, can in principle help
discriminating between Poissonian models and scenarios where the
variations are due to coherent oscillations of a single entity, but
these require more data points than are currently available from
quasar optical monitoring studies. Therefore, until proven wrong, the
Poissonian description may be regarded as a useful tool to unveil
physically meaningful properties of AGN.

\section{Summary and Conclusions}

\label{sec:Conclusions}

We have reviewed the Poissonian formalism for quasar variability in an
attempt to provide a general framework which allows fundamental
parameters to be estimated from good monitoring data. This was applied
to the 6 yr long B \& R light curves of 42 PG quasars obtained by G99,
yielding constraints for the energy, rate and lifetimes of the flares.
Our main results can be summarized as follows.

(1) The only reasonable way to account for the fact that quasars vary
    more at shorter wavelengths within a Poissonian scenario is to
    include the diluting effects of an underlying ``non-variable
    background'' component redder than the spectral energy
    distribution of the putative flares.

(2) A wide range of flare energies, lifetimes, and/or background
    fractions has to be invoked to account for the observed
    variability properties (amplitudes and time-scales). This
    ``stretching'' of the simplest Poissonian scenario (in which all
    parameters are the same for every quasar) is warranted by the
    model independent result that quasars present a wide range of
    variability time scales, from $\sim 0.5$ to more than 3 yr, as
    inferred from a Structure Function analysis.

(3) The mean number of living flares is constrained to be of order
    ${\cal N} \sim 5$ to 100, and lower limits for the R-band
    background contribution of typically 25\% are established.  These
    estimates are independent of cosmology, K-correction and
    extinction, and little sensitive to the SF analysis.

(4) Flare rates between $\sim 1$--100 yr$^{-1}$ and monochromatic
    flare energies in the $\sim 10^{46-48}$ erg$\,$\AA$^{-1}$ range
    are implied by the data. Overall, the Poissonian parameters for
    individual quasars are constrained to within about an order of
    magnitude.

(5) Light Curve simulations were performed and demonstrate the ability
    to reproduce the observed morphology of quasars light curves
    extremely well even for a blatantly simplistic ``on/off''
    square-shaped model for the evolution of the individual flares.
    Indeed, the whole Poissonian analysis is highly insensitive to the
    flare shape.

(6) Experiments were performed on the use of higher moments of the
    light curve as further constraints, but found to be of little use
    at present.

(7) The variability properties of the PG quasars present little
    correlation with other properties. Even the best correlations
    identified, like that between the variability amplitude and
    EW(H$\beta$), present a substantial scatter, confirming the
    results of G99.

(8) The EW(H$\beta$) $\times$ variability amplitude was interpreted in
    a scenario where only the variable component participates in the
    ionization of the line emitting gas, consistent with conclusion
    (1) above. Correlations with EW(HeII) and the X-ray to optical
    spectral index further support this interpretation.

Progress on this line of approach to AGN variability will require
longer light curves to obtain more accurate estimates of the
variability properties, which are important to constrain both
Poissonian and non-Poissonian models.  Regardless of possible
improvements on the time-series analysis techniques, it is important
to encourage the continuation of the current CCD monitoring campaigns.
Spectral information, some of which is already available (Kaspi \etal
2000), will also be valuable in deriving better constraints for the
model parameters and a more detailed picture of the spectral behavior
of the flares and background component.

The most appealing aspect of Poissonian models is their generality.
Scenarios as diverse as accretion disk instabilities, stellar
collisions and supernovae all fall under the large Poissonian
``umbrella''. Detailed modeling will be required to bridge the gap
between this mathematical formalism and the physics of AGN
variability and to fully explore the constraints made possible by the
application of this technique.

\acknowledgments

It is a pleasure to thank the Wise Observatory group for their
generous attitude of releasing to the scientific community the results
of a laborious and lengthy observational project. We also thank
T. Heckman and J. Krolik for their comments on an earlier version of
this paper. RCF thanks the hospitality of Johns Hopkins University,
where this work was developed. Support for this work was provided by
the National Science Foundation through grant \# GF-1001-99 from the
Association of Universities for Research in Astronomy, Inc., under NSF
cooperative agreement AST-9613615.  LV acknowledges an MSc fellowship
awarded by CAPES. Partial support from CNPq, FAPESP and PRONEX are
also acknowledged.

\appendix

\section{Skewness and Kurtosis in Poissonian models}

The Poissonian formalism allows the computation of higher moments of
the $L(t)$ distribution, though most applications to AGN variability
stop on the variance (eq.~[\ref{eq:rms_L}]). In this appendix we present
expressions for the theoretically expected third (skewness) and fourth
(kurtosis) moments, and discuss their applicability to the G99 data
set. These moments were not used to constrain the flare properties of
quasars in the main text, and are presented here only for the purposes
of completeness and future reference.

The skewness ($\gamma$) and kurtosis ($\kappa$) are defined by

\begin{equation}
\label{eq:def_skewness}
\gamma = \sigma^{-3} \ov{[L - \ov{L}]^3}
\end{equation}

\begin{equation}
\label{eq:def_kurtosis}
\kappa = \sigma^{-4} \ov{[L - \ov{L}]^4} - 3
\end{equation}

\ni The $-3$ term in (\ref{eq:def_kurtosis}) makes the kurtosis of a
Gaussian distribution 0. Expressions for $\gamma$ and $\kappa$ can be
derived extending the formalism employed by Cid Fernandes (1995),
which yields (Vieira 2000)

\begin{equation}
\label{eq:skewness}
\gamma = \phi_\gamma (\nu \tau)^{-1/2}
\end{equation}

\begin{equation}
\label{eq:kurtosis}
\kappa = \phi_{\kappa} (\nu \tau)^{-1}
\end{equation}

In these expression $\phi_\gamma$ and $\phi_\kappa$
play the role of ``shape factors'':

\begin{equation}
\label{eq:phi_skewness}
\phi_\gamma = \frac{\tau^2}{E^3} \int l^3(t) dt
\end{equation}

\begin{equation}
\label{eq:phi_kurtosis}
\phi_\kappa = \frac{\tau^3}{E^4} \int l^4(t) dt
\end{equation}

\ni Both $\phi_\gamma$ and $\phi_\kappa$ equal 1 for square shape
flares. Exponential flares result in $\phi_\gamma = 4/3$ and
$\phi_\kappa = 2$, while for triangular flares $\phi_\gamma = 9/8$ and
$\phi_\kappa = 27/20$. The above relations reveal interconnections
between the different moments in a Poissonian model: $\gamma \propto
\delta$ and $\kappa \propto \delta^2$, where $\delta = \sigma /
\ov{L}$ (eq.~[\ref{eq:rms_L}]). Unlike for $\delta$, an underlying
non-variable component does {\it not} affect $\gamma$ nor $\kappa$. In
principle, this allows an estimate of the background fraction from the
ratio of $\gamma$ (or $\kappa$) to $\delta$.

The main problem concerning the use of higher light curve moments is
that, as can be seen in eqs.~(\ref{eq:skewness}) and
(\ref{eq:kurtosis}), the predicted moments rapidly tend to the Gaussian
limit ($\gamma = \kappa = 0$) with increasing ${\cal N} = \nu \tau$,
i.e., as the superposition of events grows higher. This is obviously a
consequence of the Central Limit Theorem (e.g., Papoulis 1965).

\begin{figure*}[t]
\epsscale{0.9} 
%\plotone{fig8.eps}
\plotone{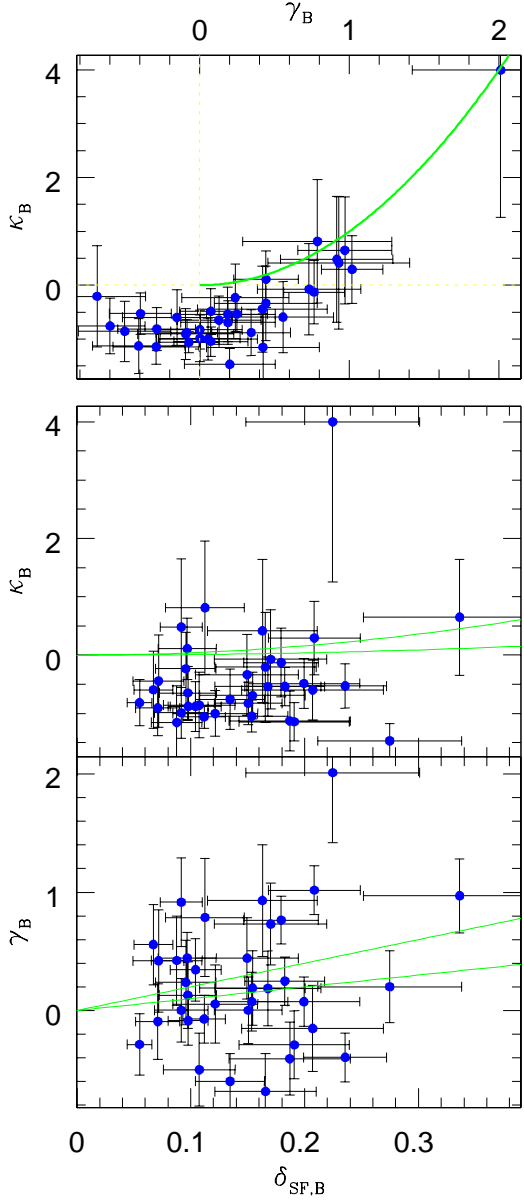}
\caption{Skewness and kurtosis for the G99 B-band light curves plotted
against each other (top panel), and $\delta_{SF}$ (bottom
panels). Error bars for $\gamma$ and $\kappa$ were computed via
bootstrap, resampling $n_{obs}$ randomly chosen points (with
repetition) from the light curve 1000 times and computing the standard
deviation of the resulting moments. The solid lines show the
theoretical predictions for $\phi_\gamma = \phi_\kappa = 1$ (square
flares). In the bottom panels the two lines correspond predictions for
a $c_B = 50\%$ background fraction (upper curves) and no background
($c_B = 0$, bottom curves). Note that $c$ affects only the abscissa in
these plots, while the $\kappa$-$\gamma$ relation is independent of
$c$.}
\label{fig:skewnes_and_kurtosis_G99}
\end{figure*}

The detection of deviations from Gaussianity in quasar light curves is
known to be problematic (Press \& Rybicki 1997), and the G99 light
curves are no exception. In Fig.~\ref{fig:skewnes_and_kurtosis_G99} we
present the skewness and kurtosis for the Wise Observatory data.
Corrections for the effects of photometric errors upon the moments
were applied, but were negligible, given the excellent accuracy of the
G99 photometry. At first sight, the results in
Fig.~\ref{fig:skewnes_and_kurtosis_G99} would seem to immediately rule
out any Poissonian model, as one finds {\it negative} $\gamma$ and
$\kappa$ for about half of the objects, whilst the theory
(eqs.~[\ref{eq:skewness}] and [\ref{eq:kurtosis}]) predict only
positive values!

%***FIG***FIG***FIG***FIG***FIG***FIG***FIG***FIG***FIG***FIG***FIG***
%***FIG***FIG***FIG***FIG***FIG***FIG***FIG***FIG***FIG***FIG***FIG***

\begin{figure*}[t]
\epsscale{1} 
%\plotone{fig9.eps}
\plotone{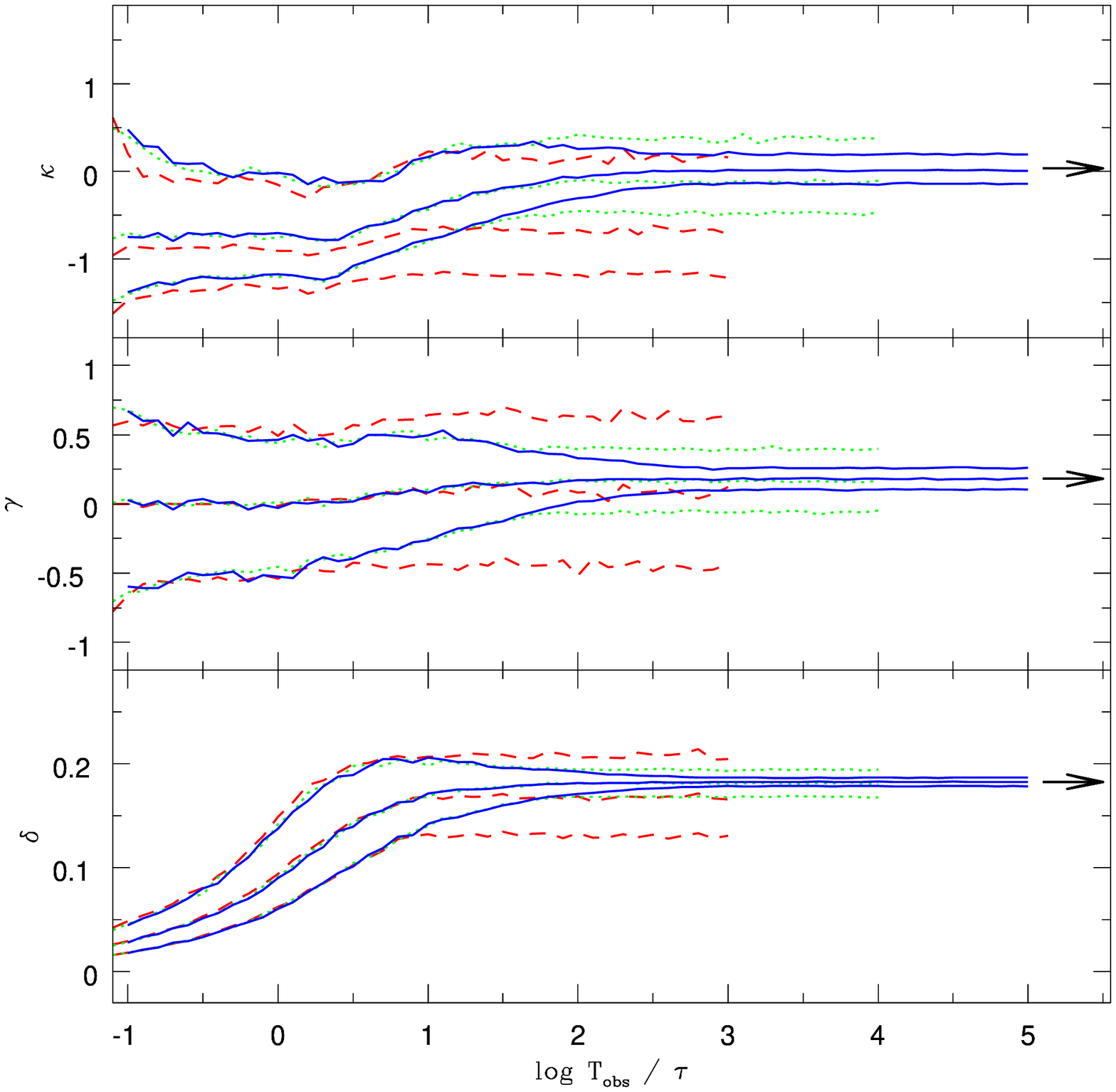}
\caption{Fractional variability ($\delta$), skewness ($\gamma$) and
kurtosis ($\kappa$) for simulations as a function of the length of the
light curve in units of $\tau$. Dashed, dotted and solid lines
correspond to $n_{obs} = 11$, 101 and 1001 observations
respectively. The middle curves indicate the median over 1000 light
curve simulations, whereas the bottom and top lines mark the 16 and
84\% percentiles respectively, such that 68\% of the points lie
between them.  Square flares and a rate of 30 events per $\tau$ were
used in the simulations. The theoretical moments are indicated by the
arrows on the right side of each panel. Large (and systematic in the
cases of $\delta$ and $\kappa$) deviations from the expected values of
the moments are observed for small $n_{obs}$ and specially for short
light curves ($T_{obs} \la 10 \tau$).}
\label{fig:Sampling_bias}
\end{figure*}

This, however, is likely to be an effect of sampling. To demonstrate
this, we have run a series of Poissonian light curve simulations and
compared the output 2nd, 3rd and 4th moments with the predicted values
as a function of observational parameters such as the length of the
light curve ($T_{obs}$) and the number of observations
($n_{obs}$). The results clearly show that negative $\gamma$ and
$\kappa$ occur very often also in simulated light curves, particularly
for $T_{obs} \la 10 \tau$, as illustrated in
Fig.~\ref{fig:Sampling_bias}. The scatter decreases for increasing
$n_{obs}$, but the agreement between analytical and simulated moments
is only achieved for large $n_{obs}$ {\it and} $T_{obs} \gg
\tau$. This {\it bias} is essentially insensitive to flare profile or
their rate. The conclusion here is that one cannot use the currently
available data to strongly constrain the higher moments of the $L(t)$
process. On the positive side, these experiments reinforce our
conclusion that the actual flare shape is irrelevant for most of the
analysis presented in this paper (see also next section).

%***FIG***FIG***FIG***FIG***FIG***FIG***FIG***FIG***FIG***FIG***FIG***
%***FIG***FIG***FIG***FIG***FIG***FIG***FIG***FIG***FIG***FIG***FIG***

We finalize by noting that, in analogy with what was done for the
second moment, higher moments should be computed from their asymptotic
($\Delta t \rightarrow \infty$) behavior estimated via SFs of the
corresponding order. These, however, are subjected to large
uncertainties due to the high powers involved and are not presented
here.

\section{Structure Functions for simple flare profiles}

The structure function of a Poissonian sequence of flares is:

\begin{equation}
SF(\Delta t) = SF(\infty) \,\, s(\Delta t)
\end{equation}

\ni where $SF(\infty) = 2 \nu E^2 / \tau$ is twice the asymptotic
variance of the $L(t)$ process, and $s(\Delta t)$, the normalized SF
of individual flares, is given by

\begin{equation}
s(\Delta t) = 1 - \frac{\int l(t+\Delta t) l(t) dt}{\int l^2(t) dt}
\end{equation}

\ni and is sensitive only to the shape of flare time profile $l(t)$.

\subsection{Square flares}

For $l(t) = l_0$ between $0 < t < T$ and 0 otherwise one obtains a a
linear SF up to $\Delta t = T$, with

\begin{equation}
s(\Delta t) = \frac{\Delta t}{T}
\end{equation}

\ni and 1 for $\Delta t > T$. The life-time $\tau$ (given by
eq.~\ref{eq:tau}) of square flares equals $T$. The dotted lines in
Fig.~\ref{fig:theoretical_SFs} show the resulting SF.

\subsection{Exponentially decaying flares}

For $l(t) = l_0 e^{-t/T}$ flares one finds

\begin{equation}
s(\Delta t) = 1 - e^{-\Delta t / T}
\end{equation}

\begin{figure*}[t]
\epsscale{1} 
%\plotone{fig10.eps}
\plotone{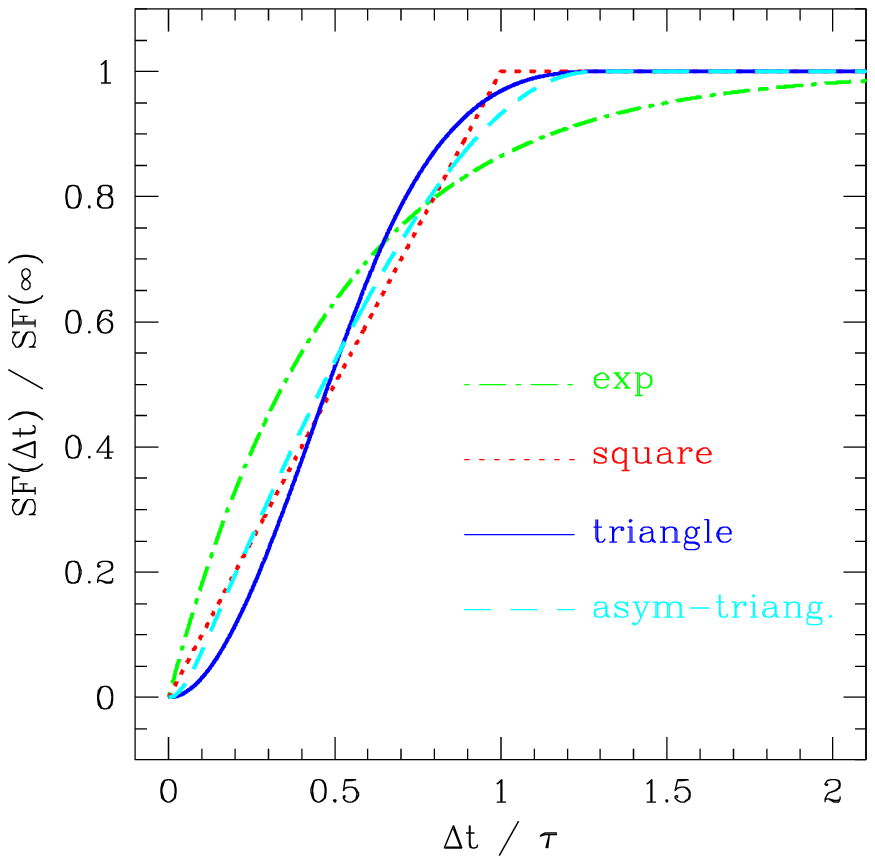}
\caption{Theoretical Structure Functions for flares with square
exponential, triangular and asymmetric triangle profiles.}
\label{fig:theoretical_SFs}
\end{figure*}

\ni while the life-time is $\tau = 2 T$. This SF is shown as a
dot-dashed line in Fig.~\ref{fig:theoretical_SFs}.

%***FIG***FIG***FIG***FIG***FIG***FIG***FIG***FIG***FIG***FIG***FIG***
%***FIG***FIG***FIG***FIG***FIG***FIG***FIG***FIG***FIG***FIG***FIG***

\subsection{Symmetric Triangular flares}

Linear flares with equal rise and decay times

\begin{eqnarray}
l(t) & = & 
  l_0 \times \left\{
  \begin{array}{lcr}
    (1 + t/T) & ; {\rm~for} & -T \le t \le 0 \\
    (1 - t/T) & ; {\rm~for} &  0 \le t \le T \\
\end{array}
\right.
\end{eqnarray}

\ni produce the following SF:

\begin{eqnarray}
s(\Delta t) = & 
  \left\{
  \begin{array}{lcr}
   -\frac{3}{4} \zp{ \frac{\Delta t}{T} }^3 
   +\frac{3}{2} \zp{ \frac{\Delta t}{T} }^2 &
      ; {\rm~for} &  0 \le \Delta t \le T \\
   +\frac{1}{4} \zp{ \frac{\Delta t}{T} }^3 
   -\frac{3}{2} \zp{ \frac{\Delta t}{T} }^2 
   + 3 \frac{\Delta t}{T} - 1 &
    ; {\rm~for} &  T \le \Delta t \le 2T \\
    1     & ; {\rm~for} &  2T \le \Delta t \\
\end{array}
\right.
\end{eqnarray}

\ni The relation between $\tau$ and $T$ for triangular shots is $\tau
= 3 T / 2$. The solid line in Fig.~\ref{fig:theoretical_SFs} shows
this SF.

\subsection{Asymmetric Triangular flares}

Triangular flares with unequal rise ($T_r$) and decay ($T_d$) times

\begin{eqnarray}
l(t) & = & 
  l_0 \times \left\{
  \begin{array}{lcr}
    (1 + t/T_r) & ; {\rm~for} & -T_r \le t \le 0 \\
    (1 - t/T_d) & ; {\rm~for} &  0 \le t \le T_d \\
\end{array}
\right.
\end{eqnarray}

\ni produce more complicated SFs:

\begin{eqnarray}
s(\Delta t) = & 
  \left\{
  \begin{array}{lcl}
	-\frac{1}{2} \frac{\Delta^3}{T}
	 \zp{ \frac{1}{T_r^2} + \frac{1}{T_r T_d} + \frac{1}{T_d^2} }
	+ \frac{3}{2} \frac{\Delta^2}{T_r T_d}
      & ; {\rm~for} &  0 \le \Delta t \le T_r \\
	1 - \frac{1}{2} \frac{\Delta^3}{T T_d^2}
	+ \frac{3}{2} \frac{\Delta}{T_d}
	- \frac{3}{2 T} 
	  \zp{ \frac{T_r^2}{6 T_d} + \frac{T_r}{2} + \frac{T_d}{3} }
      & ; {\rm~for} &  T_r \le \Delta t \le T_d \\
	-\frac{1}{2} + \frac{1}{2} \frac{\Delta^3}{T T_r T_d}
	+ \frac{3}{2} \frac{\Delta^2}{T_r T_d}
      &  \\
	+ 3 \frac{\Delta}{T} 
	  \zp{ 1 + \frac{T_d}{2 T_r} + \frac{T_r}{2 T_d} }
	- \frac{1}{2 T}
	  \zp{ \frac{T_d^2}{T_r} + \frac{T_r^2}{T_d} }
      & ; {\rm~for} &  T_d \le \Delta t \le T \\
	0
      & ; {\rm~for} &  T \le \Delta t \\
\end{array}
\right.
\end{eqnarray}

\ni where $T = T_r + T_d$. The solution above corresponds to $T_r <
T_d$. The corresponding expression for flares that spend more time
rising than decaying is identical, with $T_r$ swapped by $T_d$.  The
life-time in either case is given by $\tau = 3 T / 4$.  G99 finding
that quasars spend more time on the rise than fading gives some
motivation to using $T_r > T_d$, but the SF, being a mean of squared
differences, does distinguish between $T_r$ and $T_d$ (see Kawaguchi
\etal 1998 for techniques to explore the time-assymetry of
flares). Fig.~\ref{fig:theoretical_SFs} shows the SF for a ratio of 10
between $T_r$ and $T_d$ (or vice-versa) as a dashed line.


\begin{references}

Abraham, Z., Romero, G. E. 1999, A\&A, 344, 61

Almaini, O., Lawrence, A., Shanks, T., Edge, A., Boyle, B. J.,
Georgantopoulos, I., Gunn, K. F., Stewart, G. C., Griffiths,
R. E. 2000, MNRAS, 315, 325

Aretxaga, I., Cid Fernandes, R. \& Terlevich, R. 1997, MNRAS, 286, 271

Aretxaga, I., Terlevich, R. 1994, MNRAS, 269, 462

Ayal, S., Livio, M., Piran, T. 2000, ApJ, in press

Begelman, M., de Kool, M. 1991, in Variability of Active Galactic
Nuclei, eds.\ H. R. Miller \& P. J. Wiita (Cambridge University
Press), p.\ 198

Binette, L., Fosbury, R. A., Parker, D. 1993, PASP, 105, 1150

Bonoli, F., Braccesi, A., Federici, L., Zitelli, V., Formiggini,
L. 1979, A\&AS, 35, 391

Boorgest, U., Schramm, K. J. 1994, A\&A, 284, 764

Cardeli, J. A., Clayton, G. C., Mathis, J. S. 1989, ApJ, 345, 245

Cid Fernandes, R. 1995, PhD Thesis, University of Cambridge

Cid Fernandes, R. 1997, Rev.\ Mex.\ Astron.\ Astrof., Conf.\ Series,
4, 210

Cid Fernandes, R., Aretxaga, I. \& Terlevich, R. 1996, MNRAS, 282, 1191

Courvoisier, T.J.-L., Paltani, S. \& Walter, R., 1996, A\&A, 308, L17

Cristiani, S., Vio, R., Andreani, P. 1990, AJ, 100, 56

Cristiani, S., Trentini, S., La Franca, F., Aretxaga, I., Andreani,
P. Vio, R., Gemmo, A. 1996, A\&A, 306, 395

Cristiani, S., Trentini, S., La Franca, F., Andreani,
P. 1997, A\&A, 321, 123

Cutri, R. M., Wi\'sniewski, W. Z., Rieke, G. H., Lebofsky, M. J. 1985,
ApJ, 296, 423

Di Clemente, A., Giallongo, Natalo, G., Trevese, D., Vagnetti,
F. 1996, ApJ, 463, 466

Edelson, R. A., Krolik, J. H., Pike, G. F.1990, ApJ, 359, 86

Edelson, R. A. \etal 1996, ApJ, 470, 364

Galeev, A. A., Rosner, R., Vaiana, G. S. 1979, ApJ, 229, 318

Garcia, A., Sodr\'e, L., Jablonski, F. J. \& Terlevich, R. 1999, 
MNRAS, 309, 803

Giallongo, E., Trevese, D., Vagnetti, F. 1991, ApJ, 377, 345

Giveon, U., Maoz, D., Kaspi, S., Netzer, H., Smith, P. 1999, MNRAS,
306, 637

Haardt, F., Maraschi, L., Ghisellini, G. 1994, ApJ, 432, L95

Hawkins, M. R. S. 2000, A\&AS, 143, 465

Hook, I. M., McMahon, R. G., Boyle, B. J., Irwin, M. J. 1994, MNRAS,
268, 305

Lyutyi, V. M. 1977, SvA, 21, 655

Kaspi, S., Smith, P. S., Netzer, H., Maoz, D., Jannuzi, B. T. \&
Giveon, U. 2000, ApJ, {\it in press}

Kawaguchi, T., Mineshige, S., Umemura, M., Turner, E. L. 1998, ApJ,
504, 671

Keenan, D. W. 1978, MNRAS, 185, 389

Kinney, A. L., Bohlin, R. C., Blades, J. C., York, D. G. 1991, ApJS,
75, 645

Krolik, J. H., Horne, K. Kallman, T. R., Malkan, M. A., Edelson,
R. A., Kriss, G. R. 1991, ApJ, 371, 541

Krolik, J. H., Done, C., Madejski, G. 1993, ApJ, 402, 432

Lehto, H. J. 1989, in Proceedings of the ESLAB Symposium, eds.\ Hunt,
J. \& Bottrick, B., (Paris:ESA), p. 499

Longo, G., Vio, R., Provenzale, A., Rifatto, A. 1996, A\&A, 312, 424

Maoz, D., Smith, P. S., Januzzi, B. T., Kaspi, S., Netzer, H. 1994,
ApJ, 421, 34

Mineshige \& Shields, G. 1990, ApJ, 351, 47

Netzer, H., Sheffer, Y. 1983, MNRAS, 203, 935

Netzer, H. \etal 1996, MNRAS, 279, 429

Netzer, H., Peterson, B. M. 1997, in Astronomical Time Series, eds.\
D. Maoz, A. Sternberg \& E. M. Leibowitz, (Kluwer Academic
Publishers), p.\ 85.

Paltani, S., Walter, 1996, A\&A, 312, 55

Paltani, S., Courvoisier, T.J.-L. 1997, A\&A, 323, 717

Paltani, S., Courvoisier, T.J.-L., Walter, R. 1998, A\&A, 340, 47

Papoulis, A. 1965, Probability, random variables, and stochastic
process, (McGraw-Hill)

Peterson, B. M. 1993, PASP, 105, 247

Peterson, B. M., Ferland G. 1986, Nature, 324, 345

Pica, A. J., Smith, A. G. 1983, ApJ, 272, 11

Pica, A. J., Smith, A. G., Webb, J. R., Leacock, R. J., Clemens, S.,
Gombola, P. P. 1988, AJ, 96, 1215

Press, W. H., Rybicki, G. B. 1997, in Astronomical Time Series, eds.\
D. Maoz, A. Sternberg \& E. M. Leibowitz, (Kluwer Academic
Publishers), p.\ 61.

Schmidt, M., Green, R. E. 1983, ApJ 269, 352

Simonetti, J. H., Cordes, J. M., Heeschen, D. S. 1985, ApJ, 296, 46

Sirola, C. J., Turnshek, D. A., Weymann, R. J., Monier, E. M., Morris,
S. L., Roth, M. R., Krzeminski, W., Kunkel, W. E., Duhalde, O.,
Sheaffer, S. 1998, ApJ, 495, 659

Stern, B., Svensson, R., Sikora, M. 1991, in Variability of Active
Galactic Nuclei, eds.\ H. R. Miller \& P. J. Wiita (Cambridge
University Press), p.\ 229

Uomoto, A. K., Wills, B. J., Wills, D. 1976, AJ, 81, 905

Vieira, L. S. 2000, MSc Thesis, Universidade Federal de Santa Catarina

Vio, R., Cristiani, S., Lessi, O., Provenzale, A. 1992, ApJ, 391, 518

Warner, B. 1995, Cataclysmic Variable Stars, Cambridge Astrophysics
Series, Cambridge University Press, (Cambridge: New York)


\end{references}
\end{document}